\documentclass[%
 reprint,
 superscriptaddress,
preprintnumbers,
 amsmath,amssymb,
 aps,
 prb,
]{revtex4-1}

\usepackage{graphicx}% Include figure files
\usepackage{dcolumn}% Align table columns on decimal point
\usepackage{bm}% bold math
\usepackage[colorlinks=true]{hyperref}% add hypertext capabilities
\usepackage{subfig}
\usepackage[version=3]{mhchem}
\usepackage{txfonts}
\usepackage{caption}
\begin{document}

\preprint{}
\title{Charging of Cu atom on Mo supported thin films of ScN, MgO and NaF}
\author{P. A. \v{Z}guns}
\email{Corresponding author: pjotr.zgun@gmail.com}
\affiliation{Department of Physics and Astronomy, Uppsala University, Box 516, 75121 Uppsala, Sweden}
\affiliation{Multiscale Materials Modelling, Department of Materials Science and Engineering, KTH - Royal Institute of Technology, SE-100~44 Stockholm, Sweden}
\author{M. Wessel}
\affiliation{Faculty of Chemistry, University Duisburg-Essen, Universit\"{a}tsstr.~5, 45141 Essen, Germany}
\author{N. V. Skorodumova}
\affiliation{Department of Physics and Astronomy, Uppsala University, Box 516, 75121 Uppsala, Sweden}
\affiliation{Multiscale Materials Modelling, Department of Materials Science and Engineering, KTH - Royal Institute of Technology, SE-100~44 Stockholm, Sweden}

\begin{abstract}

Molybdenum supported thin films of ScN, MgO and NaF with a Cu adatom have been 
studied in the framework of density functional theory. We have observed a 
charge transfer from the metal/film interface to the Cu atom and investigated 
its relation to surface and interface deformations. We find that a weak 
interaction between the metal and the film is a promising prerequisite for 
adatom charging. The detailed study of Cu/NaF/Mo and NaF/Mo indicates that the 
distortion of the NaF film caused by the Cu adsorption has essentially 
anharmonic character, as it is coupled to a strong charge redistribution in 
the system.

%\begin{description}
%\pacs{68.43.Bc, 31.15.A-, 71.15.Mb, 68.55.-a}
%\item[\pacs{68.43.Bc, 31.15.A, 71.15.Mb, 68.55.-a}]
%\end{description}
\end{abstract}

%\pacs{68.43.Bc, 31.15.A, 71.15.Mb, 68.55.-a}% PACS, the Physics and Astronomy

\maketitle

%\tableofcontents

\section{\label{sec:intro}Introduction}
Modern experimental techniques made it possible to grow epitaxial films with 
atomic precision creating a whole new class of complex materials with unique 
properties where interfaces often play the decisive role \cite{SuntolaALE, 
OxSu, Fr07, FrPa08, Honk14}. Ultrathin films of insulating materials grown on 
a metal support are one example of such  materials \cite{Fr07, FrPa08, Honk14, 
MgOMo91, ReneEpit00, ScottEpit00}. In particular, due to the proximity of the 
insulator/metal interface to the vacuum/insulator interface the adsorption 
properties of such complex substrates can be completely different from those 
of the surfaces of the corresponding insulating materials \cite{Fr07, FrPa08, 
Honk14, Pa05, Gi05-72, Bro06, Ster07, Honk07, Frond08}. This makes complex 
substrates very interesting for applications in the field of heterogeneous 
catalysis \cite{Fr07, FrPa08, Honk14}.

In this regard, one of the intriguing properties of metal supported thin films 
is a charging of neutral atoms adsorbed on them. This phenomenon was first 
found when the charge of an Au atom adsorbed on NaCl supported by Cu was 
manipulated with scanning tunneling microscope (STM) \cite{Repp04}. The gold 
atom was reversibly switched between the two states: stable \ce{Au^0} and 
charged \ce{Au^-}. This experimental finding was also supported by density 
functional theory (DFT) calculations \cite{Repp04}.

A spontaneous adatom charging was for the first time predicted in DFT 
calculations for an Au atom deposited onto MgO/Mo~\cite{Pa05}. The gold atom on this 
substrate appeared to be negatively charged and its adsorption energy turned 
out to be significantly increased as compared to the adsorption of Au on MgO 
\cite{Pa05}. These theoretical findings were later supported by experimental 
results \cite{Ster07}. More recently the charging effect was intensively 
studied theoretically and experimentally for different combinations of metal 
support and thin film materials, and adsorbates (see, for example, 
Ref. \onlinecite{Gi05-72, Gi05-73, Gi06-PCCP, Gron06, Bro06, Honk07, Gi07-JCP, 
Prada08, Nil08, Hellman08, Frond08, Au-FeO-08, Simic08, Lin09, Sicolo09, 
Nil10, Mart10-CPC, He09-CO-O2, Ricci06, Ster07clust, Frond07clust, Nilius_12,
Nog2, Nog1}). 

Summarizing the current knowledge about metal supported thin films and their 
adsorption properties we notice the following facts. Thin films can 
substantially reduce the work function of support metals \cite{Gi05-72, 
Gi05-73, Gi06-PCCP, Prada08}, which should lead to easier charge transfer. 
However, the charging of adatoms can also happen despite the increase of work 
function \cite{Hellman08}. For charging adatom should have high enough 
electron affinity \cite{Pa05, Gi05-72}.

%This research was not limited only to the adsorption of 
%atoms but also included molecules, i.e. \ce{O2} \cite{He09-CO-O2} and 
%\ce{NO2} \cite{Gron06, Bro06}, as well as metal clusters 
%\cite{Ricci06, Ster07clust, Frond07clust, Frond08}. 
%It was also revealed that charging could be absent in some cases as, for 
%example, for Pd on MgO/Mo \cite{Gi05-72}, and in some systems the adatom could 
%be positively charged \cite{Gi06-PCCP, Au-FeO-08}. Further we will use the 
%word \textquotedblleft charging\textquotedblright\ to describe spontaneous 
%negative charging of an adatom if not stated otherwise.

The origin of 
adatom charge was found to be the metal and oxide interface as shown for 
Au/MgO/Ag \cite{Honk07}, \ce{NO2 \slash MgO \slash Ag} \cite{Gron06}, 
\ce{NO2 \slash BaO \slash Pt} \cite{Bro06}, and 
\ce{NO2 \slash Al2O3 \slash Ag} \cite{Hellman08}. The electron abstraction 
from the interface modifies the oxide/metal interaction and increases the 
adhesion between film and metal \cite{Gron06, Honk07}. 
It was also established that the 
charging of adatom is a long-range phenomenon and could occur for relatively 
thick films \cite{Frond08}. Honkala et al. \onlinecite{Honk07} showed that the 
charge of Au on MgO/Mo was almost independent of the thickness of MgO ranging 
from 1 to 5 monolayers.

%Also, it was found 
%that lower work function of metal supported film usually yields higher 
%adsorption energy of charged adsorbate \cite{Frond08}.

%The interplay between different contributions to the adsorption energy (e.g. 
%polarization of the film and metal, binding between the film and adatom, the 
%modification of film-metal adhesion, etc.) was studied by Frondelius et al., 
%who reported that all the contributions were intermixed and depended on the 
%charge state of the adatom \cite{Frond08}. 

%Summarizing the current knowledge about metal supported thin films and their 
%adsorption properties we notice the following facts. Thin films can 
%substantially reduce the work function of support metals \cite{Gi05-72, 
%Gi05-73, Gi06-PCCP, Prada08}, which should lead to easier charge transfer. 
%However, the charging of adatoms can also happen despite the increase of work 
%function \cite{Hellman08}. For charging adatom should have high enough 
%electron affinity \cite{Pa05, Gi05-72}.

Although the formation of negatively charged species on metal supported ultrathin films
is quite a widespread phenomenon, the mechanism behind the charging is still under debate.
First, it was suggested to be a 
direct electron tunneling from the support metal conduction band states to 
the adatom empty states~\cite{Pa05,Gi05-72,Gi06-PCCP, Mart10-CPC}. However, many authors avoid this concept 
(see, e.g. Refs.~\onlinecite{Bro06, Honk07, Frond08, Pa14, Nog2}), 
emphasizing instead the role of electrostatic interaction and system 
polarization~\cite{Bro06, Honk07, Frond08, Nog2}.
We also notice that charge redistribution at metal/insulator interfaces
was intensively studied in
semiconductor physics~\cite{Schottky38, Schottky39, Mott38, Mott39}
and a number of useful concepts, such as band bending and space charge layer formation
were developed~\cite{Schottky38, Schottky39, Mott38, Mott39, Band_Bending_Review}.

%Due to charging the adsorption energy of adatoms on metal supported films is 
%higher than that on the surfaces of the corresponding insulators 
%\cite{Pa05, Gi05-72}.
It is known that charging is accompanied by surface distortions 
around the adsorbed adatom \cite{Pa05, Gi05-72}. Moreover, calculations show 
that there is no charging on \textquotedblleft frozen\textquotedblright\ 
surfaces \cite{Gi07-JCP, Frond08}.
%Another important thing associated with adatom charging is the rumpling of the 
%film at the film/metal interface. It was shown that in some systems charging 
%is also accompanied by rumpling inversion of the oxide monolayer at the 
%interface. In particular in \ce{NO2 \slash Al2O3 \slash Ag(111)} the O atoms 
%come closer to the Ag layer, while in the absence of \ce{NO2} the Al atoms are 
%closer to the Ag layer \cite{Hellman08}.
The importance of rumpling and surface distortion
was most consistently studied by 
Goniakowski et al. (Refs.~\onlinecite{Nog2, Nog1}), who showed that the charge 
transfer from the metal into the thin film or adatom is quasilinearly 
proportional to the rumpling of the interface. 
Moreover, these works reported an important finding that the relationship between
the charge transfer and film deformation were similar for both bare supported films
and the films with adatoms~\cite{Nog2, Nog1}.
Their results suggest that interface rumpling is a response of the system
to the spontaneous charge redistribution at the interface~\cite{Nog2, Nog1}.

Here we report the results of a systematic study of the ScN/Mo, MgO/Mo and 
NaF/Mo systems with a Cu adatom. MgO, ScN and NaF were chosen for their simple 
rock-salt structure and small lattice mismatch with Mo. The anions of these 
compounds are neighbours in the Periodic Table of elements that can yield a 
trend in properties. Moreover, NaF and MgO are ionic compounds, while ScN is 
more covalent and is a semiconductor, which allows us to do a comparative 
study.
The questions we focus on are why and when charging takes place,
how it happens, which atoms and electronic states are involved in
the charging process.

This article is organized as follows. Section~\ref{sec:methodology} describes 
methodology and computational details. Section~\ref{sec:results} contains 
results and discussions. In Section~\ref{subsec:AB-Mo} and 
\ref{subsec:Cu-AB-Mo} we describe systems AB/Mo and Cu/AB/Mo 
(AB = ScN, MgO, NaF), respectively, in particular, factors enhancing charging 
as well as  connection between the charge transfer and deformation of the film 
induced by adsorption. In Section~\ref{subsec:charge_localization} we 
investigate the origin of the adatom charge and where it accumulates. Next, in 
Section~\ref{subsec:distortion_and_pumping} we demonstrate how the charge is 
\textquotedblleft pumped\textquotedblright\ from film/metal interface to the 
top surface layer, and show which electronic states are involved. In 
Section~\ref{sec:conclusions} we summarize our findings and provide concluding 
remarks.

\section{\label{sec:methodology}Methodology and computational details}

\subsection{\label{subsec:model}Model Description}

\begin{figure}
\includegraphics{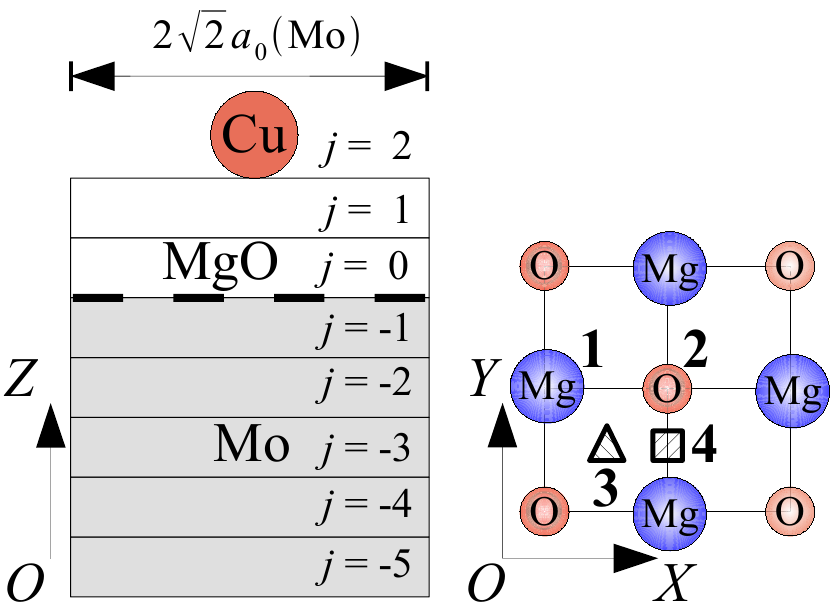}
\caption{\label{fig:slab}(Color online) Left panel: View of the upper half of 
a $2 \times 2$ symmetric Cu/MgO(2ML)/Mo(9ML)/MgO(2ML)/Cu slab. Monolayers are 
enumerated with index $j$. The dashed line represents the MgO/Mo interface and 
the middle symmetry layer of Mo has index $j = -5$. The vacuum gap between 
repeated slabs is not shown. Right panel: Top view of the MgO surface showing 
four adsorption sites: 1) Mg on-top, 2) O on-top, 3) hollow, 4) bridging.}
\end{figure}

The ultrathin films of three binary compounds on the surface of bcc molybdenum 
(AB/Mo, AB = ScN, MgO, NaF) both bare and with an adsorbed Cu atom 
(Cu/AB/Mo) have been studied. Such epitaxially grown thin films usually adopt 
rocksalt structure \cite{Rocksalt-ScN, Rocksalt-MgO}, which we consider here. 
Thus, an AB/Mo interface  is constructed of  the AB (001) surface and Mo (001) 
surface, rotated by $45^{\circ}$ with respect to each other. The anion atoms 
(B = N, O, F) are located directly above the Mo atoms, while the cation atoms 
(A = Sc, Mg, Na) are above the hollow sites of the Mo lattice. The 
calculations were carried out for the $2 \times 2$ symmetric slabs, having two 
AB monolayers (ML) on both sides of a 9 ML thick molybdenum slab 
(Fig.~\ref{fig:slab}). The thickness of the Mo slab  was chosen based on the 
convergence of the following characteristics \emph{i}) distances between 
adjacent (001) Mo planes, \emph{ii}) density of states (DOS) of surface Mo 
layer, \emph{iii}) surface energy. Also, the middle layer of the slab should 
possess bulk-like properties. Our tests show that to have a good 
representation of the mentioned properties one can use 9 ML thick Mo slabs. 
The repeated slabs were separated by at least 17 \AA\ of vacuum.

Since the films are ultrathin, while the metal support is supposed to be 
bulk-like \cite{Rocksalt-MgO} we matched the lattice parameters of the films 
to the one of Mo and aligned them with $\sqrt{2} \times a_0(\text{Mo})$  
($\sqrt{2}$ factor appears due to the Mo(001) surface rotation by 
$45^{\circ}$). The lattice mismatch parameters 
$f_m = a_f / a_s - 1$ ($a_f$ and $a_s$  are lattice constants of the film and 
support) for ScN, MgO and NaF on Mo are $1.2 \%$, $-5.3 \%$ and $4.1 \%$, 
respectively (as calculated from the experimental lattice constants 
Refs. \onlinecite{Mo-a0, ScN-a0, MgO-a0, NaF-a0}). Plus sign here means that 
the film is compressed with respect to the bulk lattice parameter, while minus 
means the film is expanded. The lattice mismatch is modest and the most 
strained film, MgO/Mo, has successfully been synthesized (see, e.g. 
Ref. \onlinecite{MgOMo91, Rocksalt-MgO}).

For all the surfaces we considered four sites for Cu adsorption: 1) on top of 
cation, 2) on top of anion, 3) hollow site above the center of the square 
comprised of two cations and two anions, and 4) bridging position above the 
center of the cation and anion bond (Fig.~\ref{fig:slab}). The Cu atom at 
these sites was allowed to relax along the $Z$ axis.

\subsection{\label{subsec:comp}Computational Details}
Density functional theory (DFT) calculations were performed using the 
projector augmented wave method \cite{PAW} together with the general gradient 
approximation (GGA) in Perdew-Budke-Ernzernhof parametrization 
\cite{PBE1, PBE2} as implemented in Vienna \emph{Ab initio} Simulation Package 
\cite{VASP1, VASP2, VASP3, VASP4, VASP5}. The cutoff energy of 800 eV was used 
in the calculations. The following states were included into the valence band: 
Cu $3d^{10}4s^{1}$, Sc $3d^{1}4s^{2}$, N $2s^{2}2p^{3}$, Mg $3s^{2}$, 
O $2s^{2}2p^{4}$, Na $2p^{6}3s^{1}$, F $2s^{2}2p^{5}$, and Mo $4d^{5}5s^{1}$. 
The Brillouin zone was sampled over a $8 \times 8 \times 1$ Pack-Monkhorst 
$k$-points mesh \cite{PM}. The calculations were spin polarized. During 
geometry optimization the relaxation of all atoms except for the Mo atoms in 
the middle symmetry layer ($j = -5$, see Fig.~\ref{fig:slab}) was allowed. A 
geometry was considered to be optimized if forces acting on unfrozen atoms 
were less than 5 meV/\AA.

We notice that even though standard GGA functionals underestimate band gaps,
they adequately describe charging phenomena in metal supported thin films
that has been demonstrated by numerous previous works \cite{Pa05,
Gi05-72, Bro06, Ster07, Honk07, Frond08, Gi06-PCCP, Gron06,
Bro06, Honk07, Gi07-JCP, Nil08, Hellman08, Frond08,
Simic08, Lin09, Sicolo09, Nil10, Mart10-CPC, He09-CO-O2, Ricci06,
Ster07clust, Frond07clust, He09-CO-O2}.
Moreover, it has recently been shown that both standard GGA and hybrid
functional produce very similar Bader charges of an Au adatom 
on MgO and CaO substrates doped with Mo or Cr \cite{hybrid2013}.

To understand the mechanism of adatom charging on the AB/Mo surfaces we 
analysed the atomic charges calculated using the Bader approach 
\cite{Bader, Bader-TEX}. We also found it elucidative to analyze the sum of 
atomic charges in separate layers (an approach also used in 
Ref. \onlinecite{Goniak04}). In order to choose an optimal grid we tested the 
convergence of Bader charges with respect to the grid density. Our tests 
showed that the Bader charge of a monolayer was much more sensitive to the 
grid step in the $Z$ direction than in the other directions that allowed us to 
save computational resources by choosing a non-uniform grid with smaller steps 
along the $Z$ axis. The step values: $\Delta z < 0.02~\AA$ and 
$\Delta x = \Delta y < 0.05~\AA$  resulted in the Bader charges with the 
accuracy of about $0.01-0.05e$ per monolayer and about $0.01e$ per atom.

\section{\label{sec:results}Results and Discussions}

\subsection{\label{subsec:AB-Mo}ScN/Mo, MgO/Mo, NaF/Mo}

First, we calculated the bulk and surface properties of the involved 
materials. The calculated lattice constant of bcc Mo, $a_0 = 3.151\,\AA$, is 
in good agreement with the experimental value of $3.147\,\AA$ \cite{Mo-a0}, 
and the calculated surface energy of Mo(001), $\mathrm{3260\, mJ/m^2}$ 
($\mathrm{0.203\, eV/\AA^2}$), is close to experimental 
$\mathrm{2930\,mJ/m^2}$ ($\mathrm{0.183\,eV/\AA^2}$) \cite{Mo_001}. The 
calculated work function of Mo(001), $\mathrm{\Phi = 3.76~eV}$, is 
underestimated compared to the experimental value of 4.53 eV \cite{Mo-WF}. The 
underestimation of the work function by GGA for similar systems is well-known 
\cite{Gi05-72}. The calculated lattice constants of bulk ScN, MgO and NaF are 
4.500 \AA\ (4.501 \AA\ \cite{ScN-a0}), 4.240 \AA\ (4.214 \AA\ \cite{MgO-a0}), 
4.705 \AA\ (4.634 \AA\ \cite{NaF-a0}), respectively (experimental values are 
given in parentheses). Notice that in the calculations of the complex 
substrates reported here the Mo lattice parameter was used. The calculated 
band gap values for ScN, MgO and NaF are 0.1 eV (1.3 eV \cite{ScN-Eg}), 
4.6 eV (7.7 eV \cite{MgO-Eg}) and 6.1 eV (11.5 eV \cite{NaF-Eg}), respectively 
(experimental values are given in parentheses). As expected the band gap 
values are underestimated.

\begin{figure*}
\subfloat[ScN/Mo]{\includegraphics[width = 0.32\textwidth]{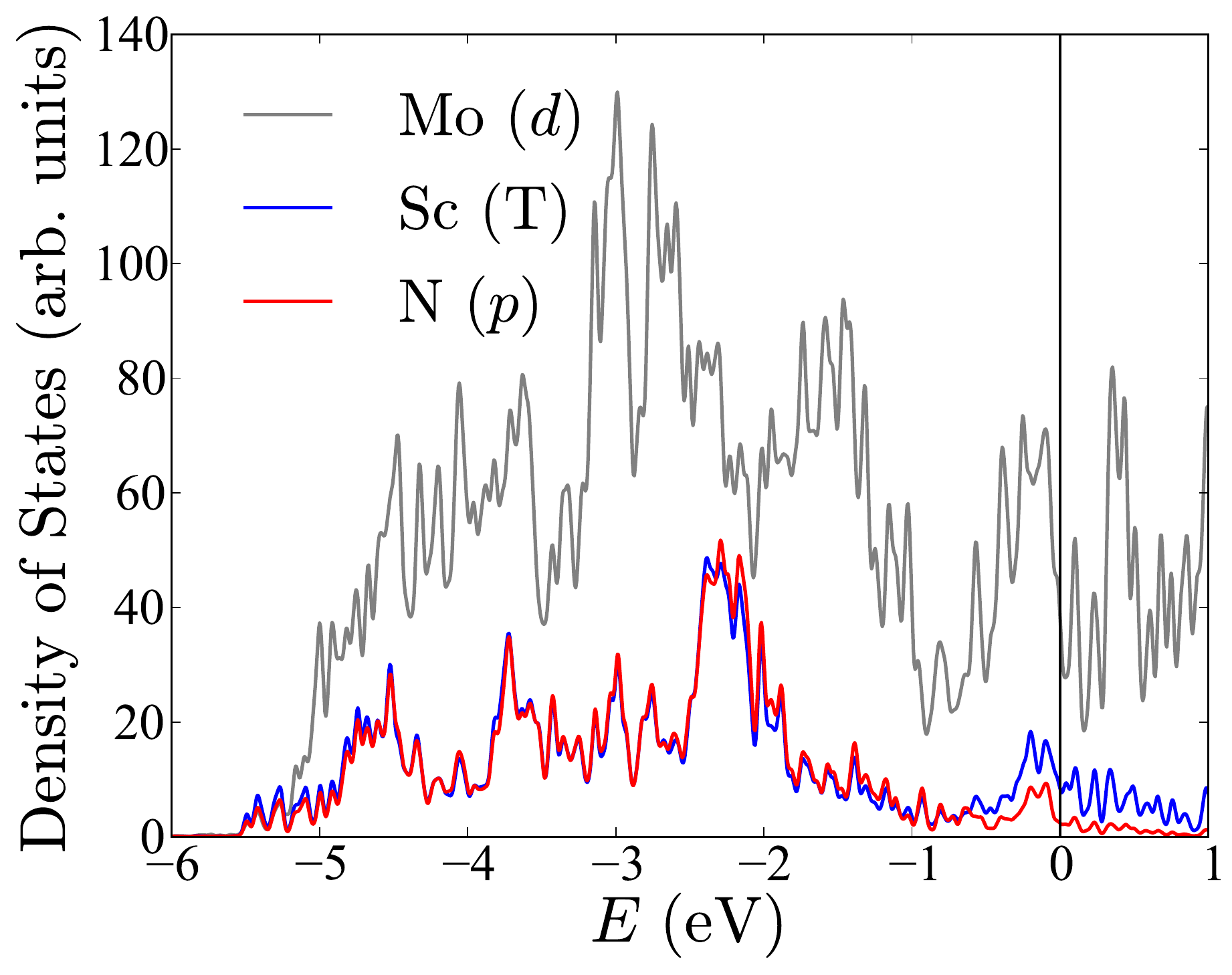}}\hfill
\subfloat[MgO/Mo]{\includegraphics[width = 0.32\textwidth]{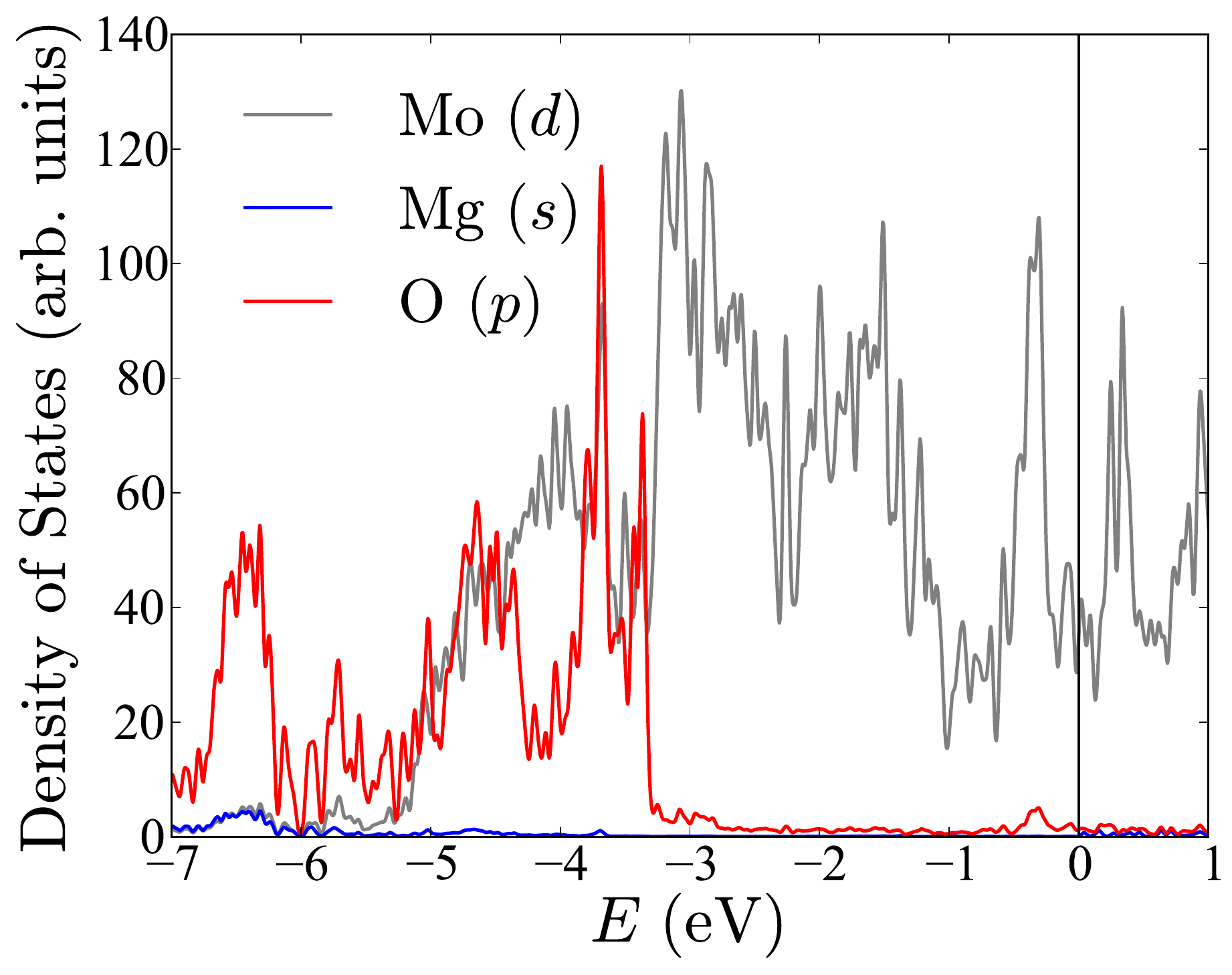}}\hfill
\subfloat[NaF/Mo]{\includegraphics[width = 0.32\textwidth]{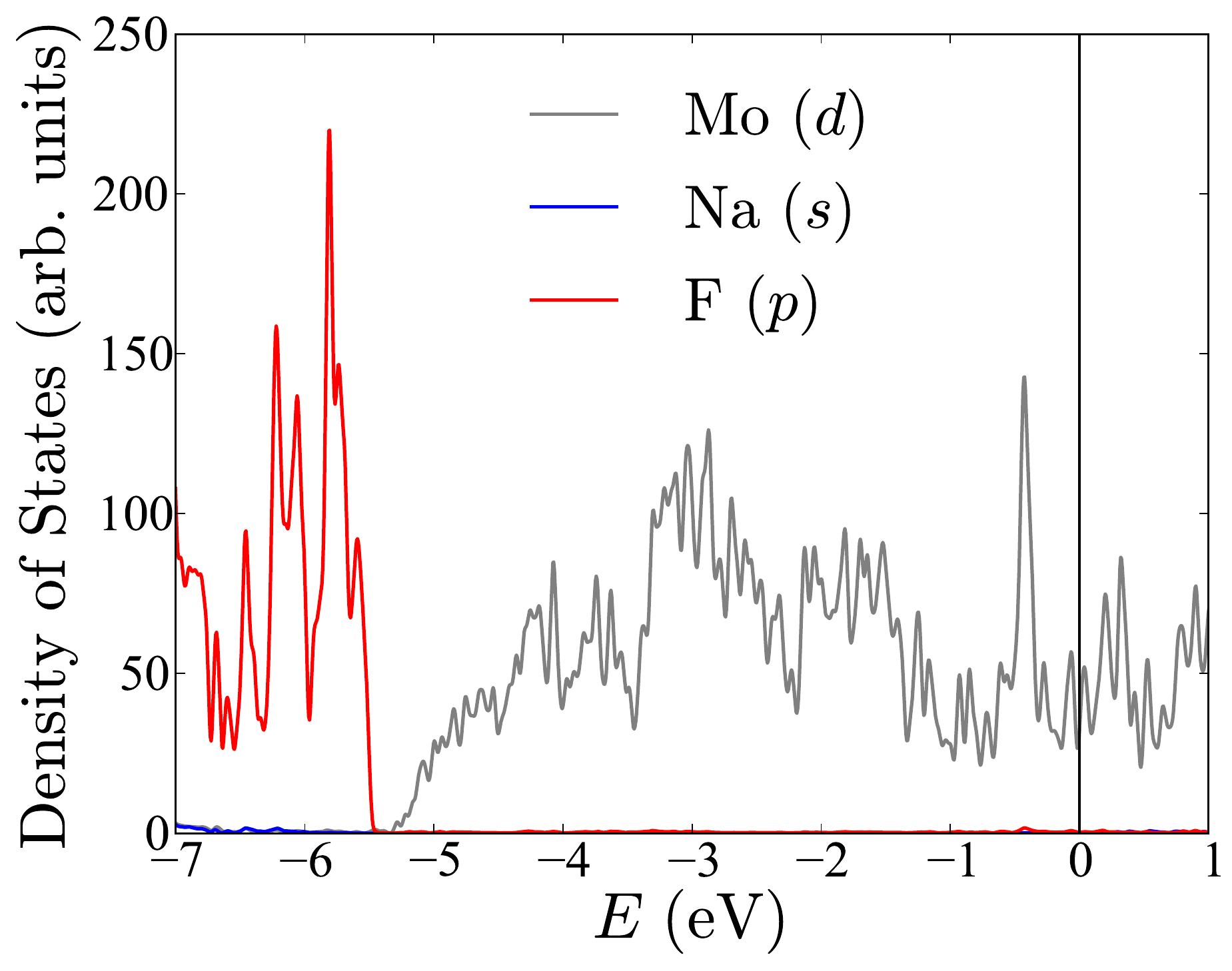}}
\caption{\label{fig:dos1}(Color online) The partial density of states of 
(a) ScN/Mo, (b) MgO/Mo, and (c) NaF/Mo. The Fermi level is set to zero. 
From left (ScN/Mo) to right (NaF/Mo) the overlap of metal and film states 
decreases drastically. For Sc total DOS (T) is shown.}
\end{figure*}

\begin{figure*}
\subfloat[ScN/Mo]{\includegraphics[width = 0.32\textwidth]{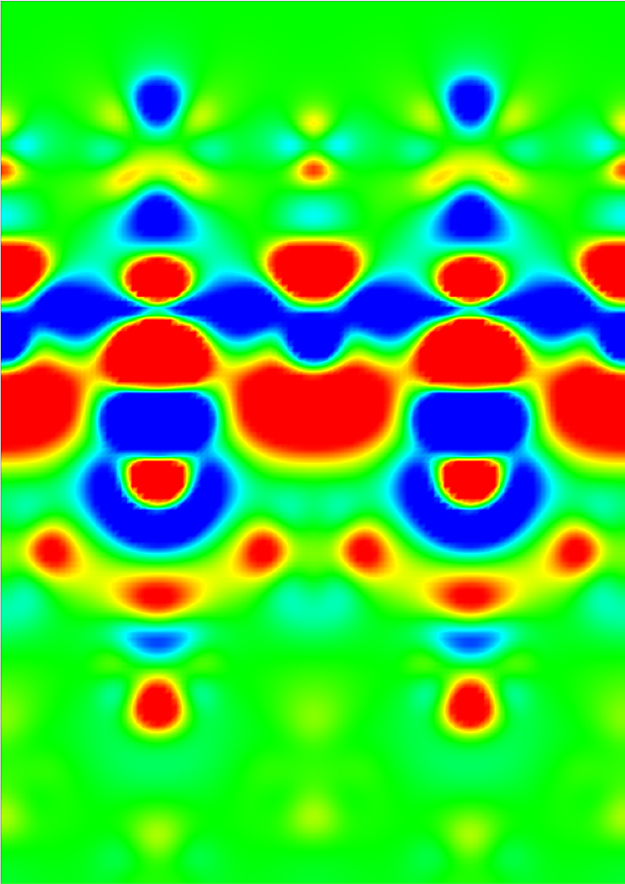}}\hfill
\subfloat[MgO/Mo]{\includegraphics[width = 0.32\textwidth]{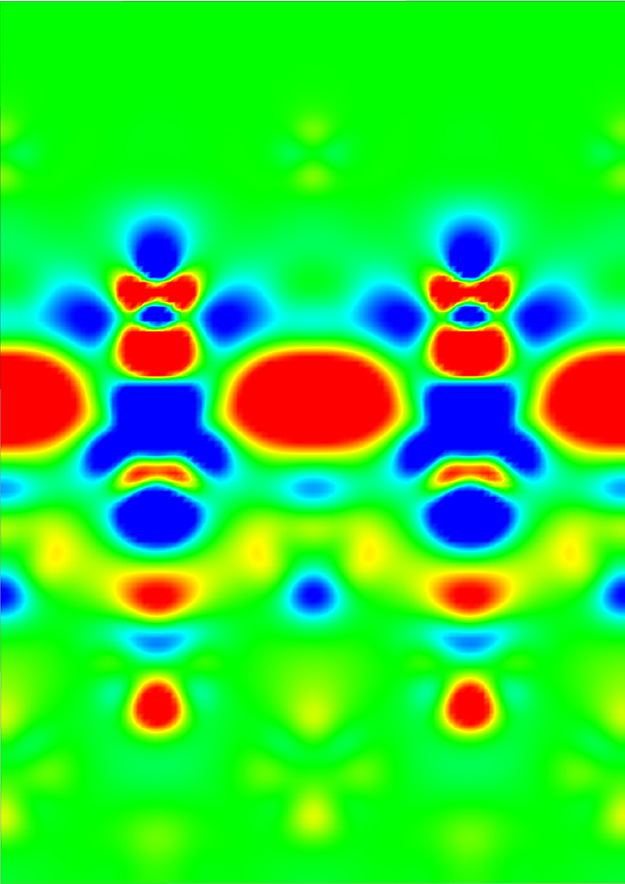}}\hfill
\subfloat[NaF/Mo]{\includegraphics[width = 0.32\textwidth]{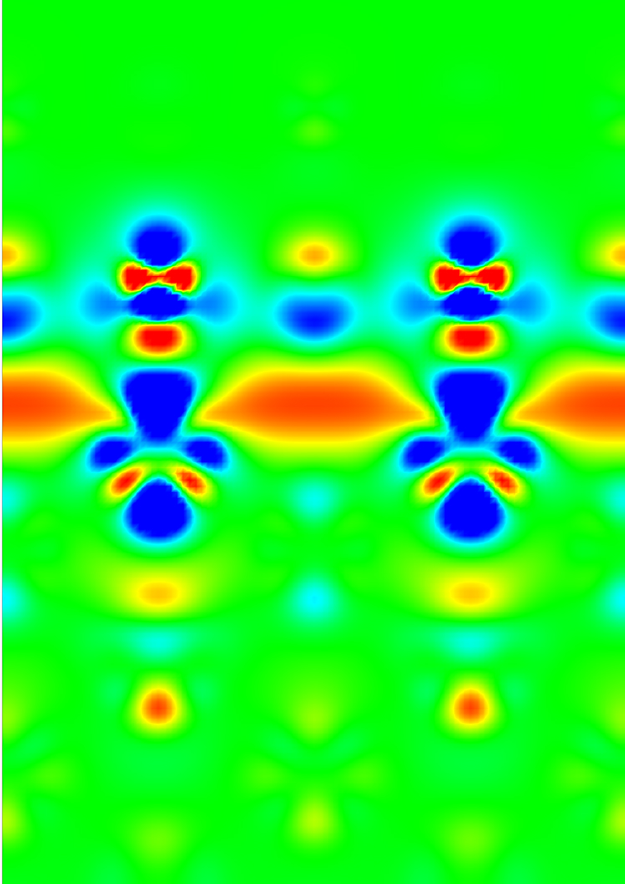}}
\caption{\label{fig:diff}(Color online). Charge density difference maps of 
$\Delta \rho = \rho(\text{AB/Mo}) - \rho(\text{AB}) - \rho(\text{Mo})$ shown 
for (100) plane: (a) ScN/Mo, (b) MgO/Mo, (c) NaF/Mo. Here, red represents an 
increase and blue a decrease of charge density. The intensity of charge 
redistribution decreases from left (ScN/Mo) to right (NaF/Mo).}
\end{figure*}

The electronic structure, charge density redistribution pattern and other 
calculated characteristics of the metal supported thin films are shown in 
Figs.~\ref{fig:dos1}, \ref{fig:diff} and Table~\ref{tab:abmo}. The densities 
of states demonstrate that the states of ScN and Mo are strongly mixed, the 
states of MgO and Mo are moderately mixed, while for those of NaF and Mo 
virtually there is no mixing (Fig.~\ref{fig:dos1}). The calculated adhesion 
energy follows this trend decreasing from ScN to NaF. The interface distance, 
on the contrary, increases from ScN to NaF (Table~\ref{tab:abmo}), indicating 
a somewhat stronger bonding between ScN and Mo compared to the other 
considered cases. The charge density difference maps (Fig.~\ref{fig:diff}) 
show a noticeable charge redistribution at the ScN/Mo interface involving even 
the second layer of ScN. In the case of MgO/Mo the charge redistribution at 
the interface is less pronounced with mostly O atoms being affected. In the 
case of NaF/Mo the redistribution is quite weak. Rumpling at the interface is 
more pronounced for MgO/Mo and NaF/Mo, than for ScN/Mo. The reduction of the 
work function is least pronounced for ScN/Mo and most pronounced for NaF/Mo.

\begin{table}
\begin{center}
\caption{\label{tab:abmo}Adhesion energy $E_{\text{adh}}$, work function 
$\Phi$, interface distance 
$d = \frac{1}{2} z_0(\ce{A}) + \frac{1}{2} z_0(\ce{B}) - z_{-1}(\ce{Mo})$, 
rumpling in the layer $r_j = z_j(\ce{B}) - z_j(\ce{A})$.}
  \begin{tabular}{l c c c}
    \hline
    & \text{ScN/MO} & \text{MgO/Mo} & \text{NaF/Mo} \\
    \hline   
	$E_{\text{adh}}~(\text{eV}/\AA^2)$  & 0.204 & 0.130 & 0.039 \\
	$\Phi$~(eV)  & 3.14  & 1.85  & 1.28 \\
	$d~ (\AA)$   & 2.11  & 2.15  & 2.41 \\
	$r_1~ (\AA)$ & 0.02  & 0.04  & 0.02 \\
	$r_0~ (\AA)$ & 0.02  & 0.08  & 0.04 \\
	\hline
	\hline
  \end{tabular}
\end{center}
\end{table}

Thus, our results demonstrate that the strength of Mo and film bonding decreases 
in the row: ScN, MgO, and  NaF (Fig.~\ref{fig:dos1}, Fig.~\ref{fig:diff}, 
and Table~\ref{tab:abmo}). Among the considered compounds ScN is the least 
ionic and NaF is the most ionic, which is also shown by the analysis of the 
Bader charges of the bulk compounds: $Q(\ce{N}) = -1.60 e$, $Q(\ce{O}) = -1.65 e$, and 
$Q(\ce{F}) = -0.87 e$. Thus we observe the following trend: for more ionic 
films we find weaker adhesion to the metal, smaller overlap of the densities 
of states of the metal and the film, and less pronounced charge redistribution 
at the interface. As we show further, the strength of the interface bonding 
is also related to the charging of an adatom on the AB/Mo surface.

\subsection{\label{subsec:Cu-AB-Mo}Cu Atom Adsorption on Mo Supported Thin Films}

Spontaneous adatom charging is studied here by considering the adsorption of 
Cu atom on the AB/Mo surfaces. Before going into details, we want to outline 
that the capability to negatively charge the Cu adatom increases in the 
sequence ScN/Mo, MgO/Mo, NaF/Mo~--- the same sequence in which the bonding 
strength, adhesion energy, intensity of charge redistribution and the work 
function of the AB/Mo substrate decrease.

In order to characterize the adsorption of Cu we investigated four adsorption 
sites: on top of cation, on top of anion, hollow and bridging, and calculated 
several quantities, such as, adsorption energy $E_\text{ads}$, Bader charge 
of Cu adatom $Q(\ce{Cu})$, the vertical distance from Cu to the top layer 
(average position of atoms) $z$, and rumpling of atomic layers 
$r_j = \langle z_j(\text{B}) \rangle - \langle z_j(\text{A}) \rangle$, 
calculated from average $z$ coordinates of A and B in the layer with an 
index $j$ (see Fig.~\ref{fig:slab}). The obtained results are summarized in 
Table~\ref{tab:cuabmo}.

\begin{table}
\begin{center}
\caption{\label{tab:cuabmo}Cu adsorption on AB/Mo (AB = ScN, MgO, NaF) 
considering four different adsorption positions. Adsorption energy 
$E_\text{ads}$, Bader charge of Cu adatom $Q(\ce{Cu})$, vertical distance $z$ 
from Cu to the top layer (average position of atoms). The average rumpling 
$r_j$, is calculated from the average $z$ coordinates of A and B in the layer. 
The adsorption on frozen surfaces (only Cu atom position was optimized) is
marked with stars (*). Bridging position on the ScN/Mo substrate was found to 
be unstable and was not  converged.}
  \begin{tabular}{l c c c c c}
    \hline
    & $E_\text{ads}~(\text{eV})$ & $Q(\ce{Cu})~(e)$ & $z~(\AA)$ & $r_0~(\AA)$ & $r_1~(\AA)$ \\
    \hline   
    ScN/Mo \\
    \hline
    Sc on-top & $0.752$ & $-0.51$ & $2.84$ & $-0.02$ & $-0.06$ \\
    N on-top  & $1.609$ & $-0.01$ & $1.96$ &  $0.04$ &  $0.03$ \\
    N on-top* & $1.496$ & $-0.07$ & $1.93$ &  $0.02$ &  $0.02$ \\
    hollow    & $0.909$ & $-0.39$ & $2.36$ & $-0.01$ & $-0.03$ \\
    \hline
    MgO/Mo \\
    \hline
    Mg on-top & $1.386$ & $-0.64$ & $2.77$ & $0.00$ & $-0.07$ \\
    O on-top  & $1.326$ & $-0.59$ & $2.26$ & $0.01$ & $-0.07$ \\
    hollow    & $1.466$ & $-0.67$ & $2.45$ & $0.00$ & $-0.07$ \\
    hollow*   & $0.848$ & $-0.32$ & $2.30$ & $0.08$ &  $0.04$ \\
    bridging  & $1.401$ & $-0.65$ & $2.50$ & $0.00$ & $-0.07$ \\
	\hline
	NaF-Mo \\
	\hline
	Na on-top & $1.651$ & $-0.76$ & $2.78$ & $-0.22$ & $-0.34$ \\
	F on-top  & $1.877$ & $-0.76$ & $2.69$ & $-0.22$ & $-0.34$ \\
	F on-top* & $0.509$ & $-0.22$ & $2.47$ &  $0.04$ &  $0.02$ \\
	hollow    & $1.740$ & $-0.76$ & $2.92$ & $-0.21$ & $-0.34$ \\
	bridging  & $1.764$ & $-0.76$ & $2.80$ & $-0.21$ & $-0.34$ \\
	\hline
	\hline
  \end{tabular}
\end{center}
\end{table}

\begin{figure*}
\subfloat{\includegraphics[width = 0.32\textwidth]{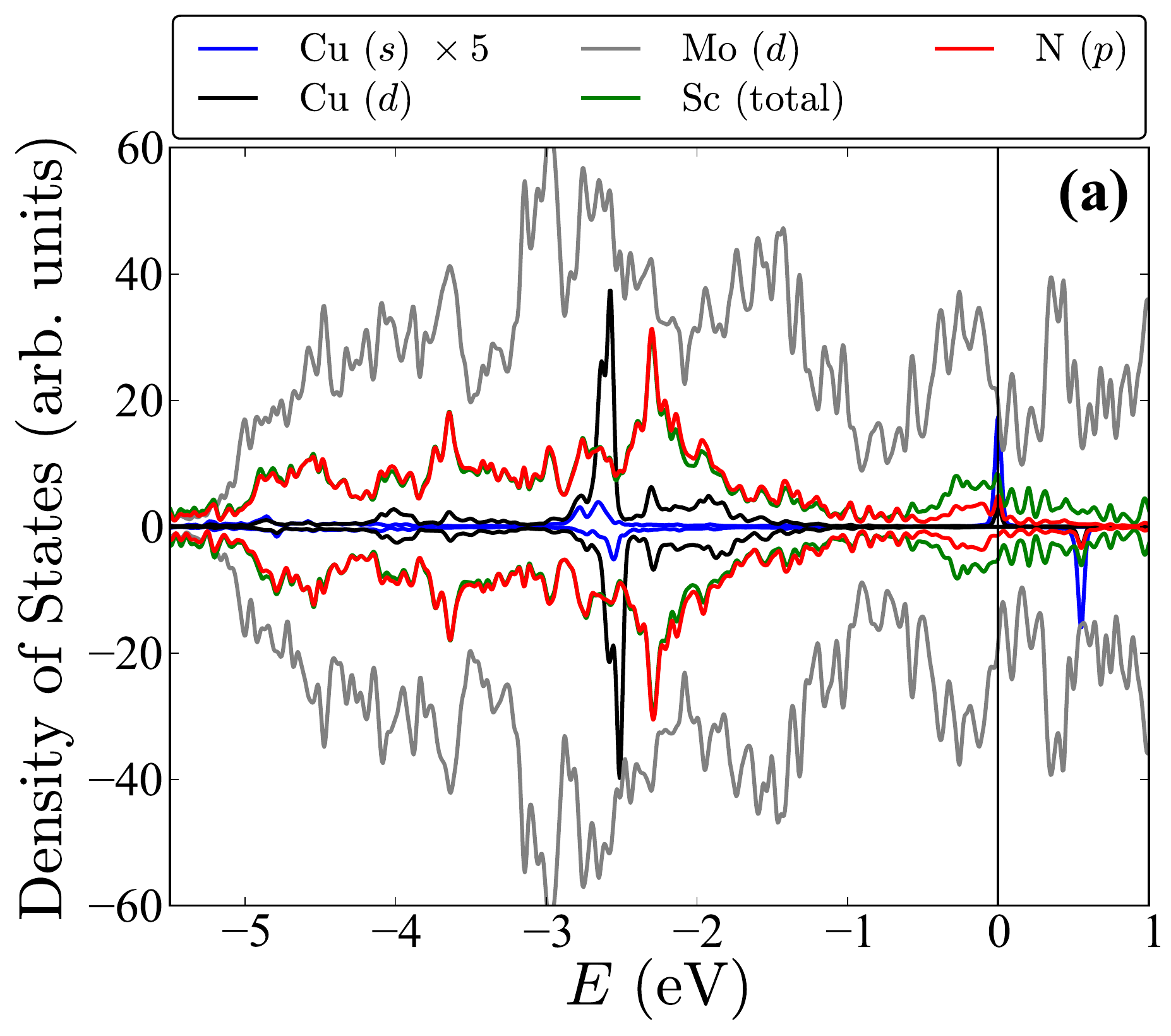}}\hfill
\subfloat{\includegraphics[width = 0.32\textwidth]{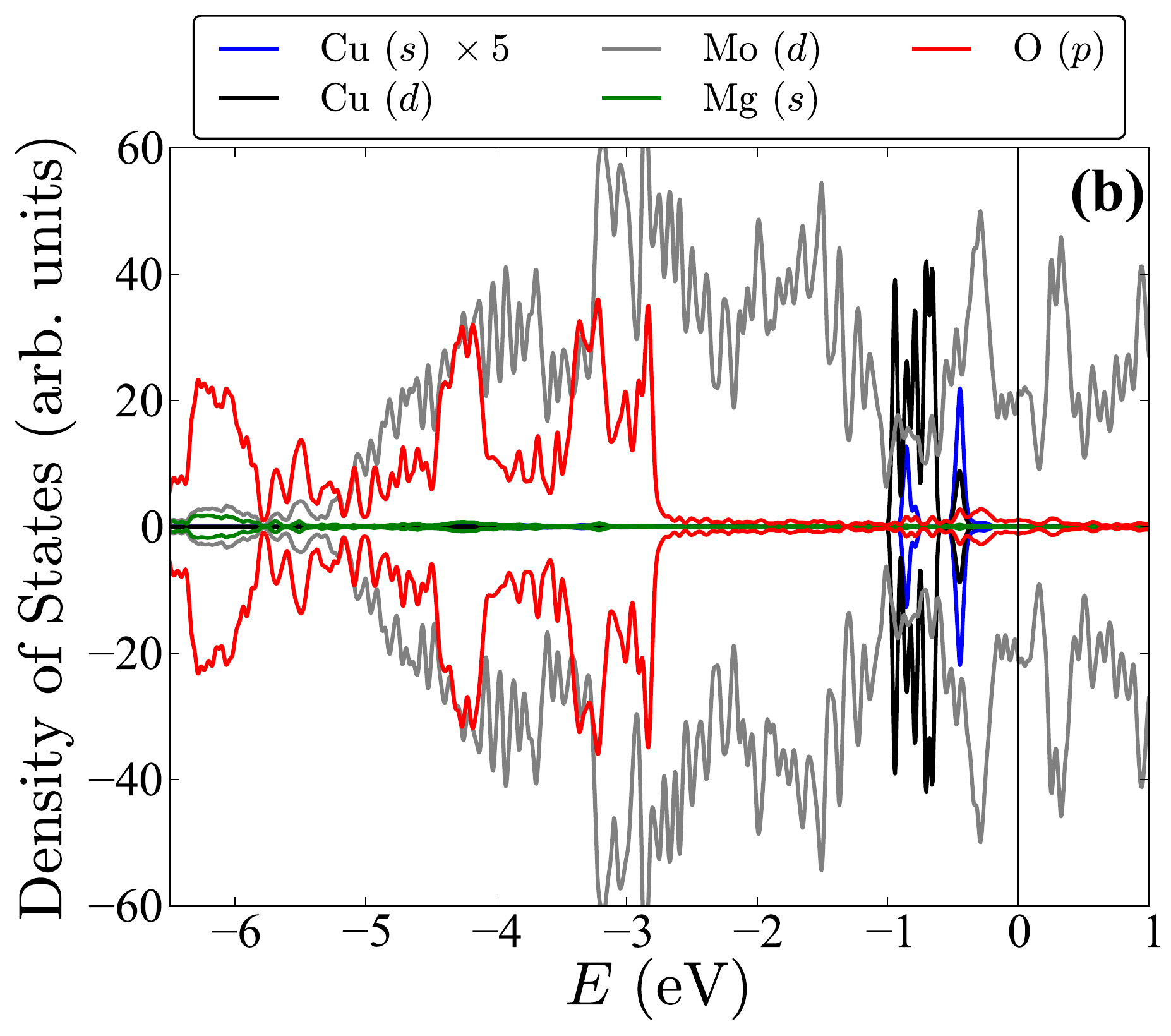}}\hfill
\subfloat{\includegraphics[width = 0.32\textwidth]{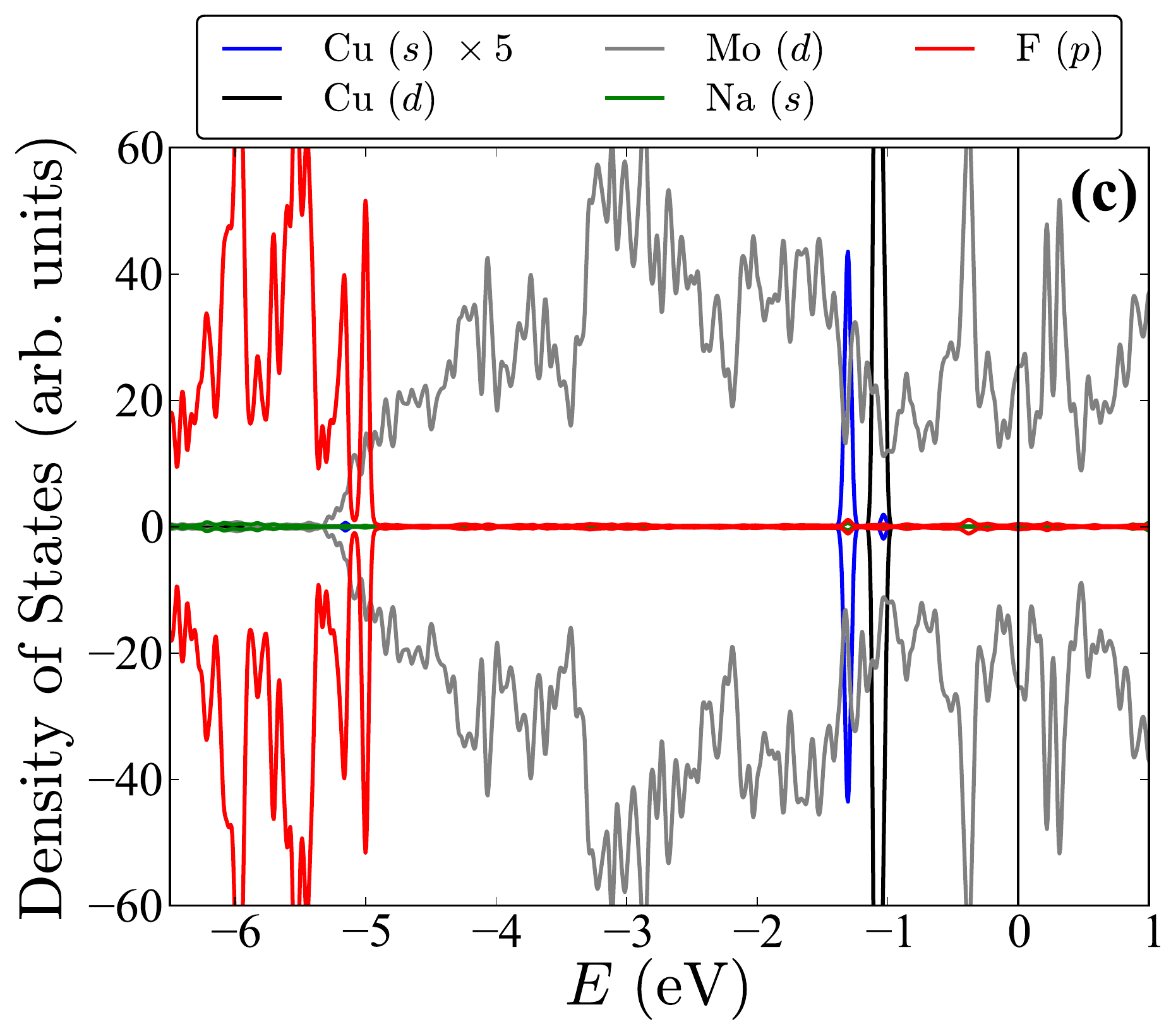}}
\caption{\label{fig:Cu_DOS}(Color online). The density of states of 
(a) N on-top site Cu/ScN/Mo, 
(b) hollow site Cu/MgO/Mo, and 
(c) F on-top site Cu/NaF/Mo. The Fermi level is set to zero.}
\end{figure*}

As one can see charging of Cu is most pronounced on NaF/Mo. For all the 
considered adsorption positions, the Cu $4s$ states fall below the Fermi 
level and no spin-polarization is observed (see the example of F on-top 
site, Fig.~\ref{fig:Cu_DOS}c). The Bader charge of Cu is almost the same 
for all the positions, $Q(\ce{Cu}) \approx -0.76 e$, which is in agreement 
with charges reported for similar systems, e.g. Au on MgO/Mo \cite{Honk07}. 
Interestingly, despite fluorine being the most electronegative element of the 
Periodic Table Cu charging occurs even for Cu adsorbed on-top of F, which was 
found to be the most favorable adsorption site.

Charging of Cu on NaF/Mo is accompanied by substantial atomic displacements in 
the top NaF layer that results in rather large adsorption energies (from 1.651 
to 1.877~eV). In particular, the F atom, situated below adsorbed Cu, moves 
downwards by 0.27~\AA , while the four nearest Na atoms move upwards by 
0.57~\AA . A similar distortion pattern was reported for Au/MgO/Mo 
\cite{Pa05}. Together with distortions around Cu we also observe a prominent 
average rumpling ($\approx -0.34~\AA$) of the NaF surface layer ($j = 1$, see 
Fig.~\ref{fig:slab}) for all the adsorption sites 
(see Table~\ref{tab:cuabmo}). The significant changes occur also at the 
interface, where the inversion of rumpling is observed and 
$z_0(\ce{F}) - z_{-1}(\ce{Mo})$ becomes smaller than 
$z_0(\ce{Na}) - z_{-1}(\ce{Mo})$. Thus upon Cu adsorption on-top of F the Mo-F 
distance on average shortens by $\sim 0.1~\AA$, as compared to that in NaF/Mo. 
For all the adsorption sites on NaF/Mo the same inversion of rumpling occurs 
(see Table~\ref{tab:cuabmo}). Similar structural changes were also reported 
for the \ce{NO2} adsorption on \ce{Al2O3 \slash Ag} \cite{Hellman08}.

The characteristics of Cu adsorption on MgO/Mo are similar to those of 
Cu/NaF/Mo. Cu charging occurs for all the four adsorption sites. The Bader 
charges of Cu, $Q(\ce{Cu})$, are smaller than those for Cu on NaF/Mo, but 
still significant ranging from $-0.59e$ to $-0.67e$. The adsorption energies 
are rather large (from 1.326 eV to 1.466 eV), the favored adsorption site is 
the hollow site. The Cu adsorption on MgO/Mo is accompanied by notable surface 
relaxation. The two Mg atoms, nearest to Cu adsorbed at the hollow site, shift 
upwards by 0.19~\AA, while two nearest O atoms move downwards by 0.11~\AA. The 
average rumpling of the surface monolayer ($j = 1$) is about $-0.07~\AA$. At 
the same time, rumpling of the interface MgO layer ($j = 0$) is negligible as 
compared to that in the Cu/NaF/Mo system. The average vertical distances from 
O ($j = 0$) to the Mo ($j = -1$) layer and from Mg ($j = 0$) to the Mo 
($j = -1$) layer are the same, 2.15 \AA . Due to the Cu adsorption the Mo-O 
distance shortens by $\sim 0.04~\AA$.

According to Bader analysis the Cu atom on MgO/Mo has a smaller charge than 
that on NaF/Mo, however, in both cases the Cu $4s$ states are filled and no 
spin-polarization is observed (Fig.~\ref{fig:Cu_DOS}). Although the Bader 
analysis cannot be expected to unambiguously assign charges to atoms in a 
solid somewhat smaller charge of Cu on MgO/Mo corresponds to a smaller 
adsorption energy and smaller displacements of surface atoms in comparison 
with those for NaF/Mo.

On ScN/Mo, Cu prefers to adsorb on top of N that yields no charging. The 
Cu $4s$ states are half occupied and mixed with the N states 
(Fig.~\ref{fig:Cu_DOS}a). Adsorption at this site results in Cu-N covalent 
bonding with the Cu-N distance of 1.86 \AA\ and the adsorption energy of 
1.609 eV. Surface relaxation in this case is very weak in comparison with 
all the cases where charging is observed.

As a matter of fact, charging of Cu occurs also on ScN/Mo for two adsorption 
sites: Sc on-top and hollow. However, even in these cases it is less 
pronounced than in the Cu/MgO/Mo and Cu/NaF/Mo systems. For Cu adsorbed at the 
Sc on-top site the Cu $4s$ states are fully occupied (not shown) and at the 
hollow site Cu $4s$ states are preponderantly occupied (not shown). The 
calculated Bader charges are still notable, $Q(\ce{Cu}) = -0.51e$ and $-0.39e$ 
for the Sc on-top and hollow sites, respectively. The interface and surface 
rumpling is also negative as in the case of Cu/NaF/Mo but somewhat less 
pronounced (see Table~\ref{tab:cuabmo}).

We notice that as Cu gets charged at all adsorption sites on both MgO/Mo and 
NaF/Mo the resulting adsorption energies are quite similar that provides a 
flat energy profile for Cu on these surfaces, especially on MgO/Mo, and, 
therefore, simplifies surface diffusion. This is not the case for ScN/Mo where 
the adsorption energies are very different, e.g. 0.752~eV and 1.609~eV for the 
Sc on-top and N on-top positions, respectively.

It is well-known that without metal support adatom charging is negligible 
(see, e.g. Ref. \onlinecite{Pa05, Amft10}). We have examined Cu adsorption 
on top of anions in the Cu/AB systems (2ML AB with the same lattice parameter 
as in the other calculations). For the N on-top site of Cu/ScN we got 
$E_\text{ads}$ = 1.523 eV and $Q(\ce{Cu}) = -0.13e$. For the O on-top site of 
Cu/MgO we got $E_\text{ads}$ = 1.110 eV and $Q(\ce{Cu}) = -0.16e$. For the F 
on-top site of Cu/NaF we got $E_\text{ads}$ = 0.380 eV and 
$Q(\ce{Cu}) = -0.12e$. Thus, in all these cases charging is negligible.

The comparison of Cu on ScN/Mo, MgO/Mo and NaF/Mo allows us to make some 
general conclusions regarding the capability of metal supported thin films to 
charge adatoms. As follows from our results this capability is enhanced from 
ScN/Mo to MgO/Mo and further to NaF/Mo. ScN is a covalent compound also 
forming covalent-like bonds with Mo, while MgO and NaF are ionic 
(see Section~\ref{subsec:AB-Mo}). Therefore, to construct a system where a 
substantial charge transfer is possible one could use ionic films whose states 
are weakly mixed with those of the metal support. Strong adatom-substrate 
bonds can also prohibit the charge transfer whereas a weaker interaction 
between the adatom and a film can promote it. Notice that Cu and MgO (NaF) 
states almost do not mix. The same trend we find for Cu adsorbed at the Sc 
on-top and hollow sites on ScN/Mo where a charge transfer takes place. Thus, 
a minimum mixing between adatom and thin film states as well as between those 
of the insulating film and the metal support appear to be crucial for charging 
to occur.

We also investigated the adsorption of Cu on frozen surfaces for the most 
stable positions: N on-top site for Cu/ScN/Mo, hollow site for Cu/MgO/Mo, and 
F on-top site for Cu/NaF/Mo (see Table~\ref{tab:cuabmo}). For the F on-top 
(Cu/NaF/Mo) and hollow sites (Cu/MgO/Mo) the adsorption energies and Bader 
charges of Cu are reduced significantly. According to the density of states 
calculated for these cases the Cu spin-down $4s$ states are split from the 
spin-up states and are not fully occupied. These findings agree with the 
results reported for the Au adsorption on frozen MgO/Ag 
\cite{Gi07-JCP, Frond08}. We also notice that in the case of Cu adsorption at 
the N on-top site of ScN/Mo the adsorption energy, Bader charge and the Cu-N 
distance calculated for the frozen and fully relaxed surfaces are very similar 
(see Table~\ref{tab:cuabmo}). This is due to the absence of charging in both 
cases, when the Cu-N bond is a dominating contribution to the adsorption 
energy. Also we observe, that if there is no charging or it is significantly 
reduced (as on frozen surfaces) the vertical distance from the Cu to the 
surface is shorter, while charged Cu shifts further away from the surface 
(see Table~\ref{tab:cuabmo}).

\begin{figure}
\centering
\includegraphics[width = 3.375in]{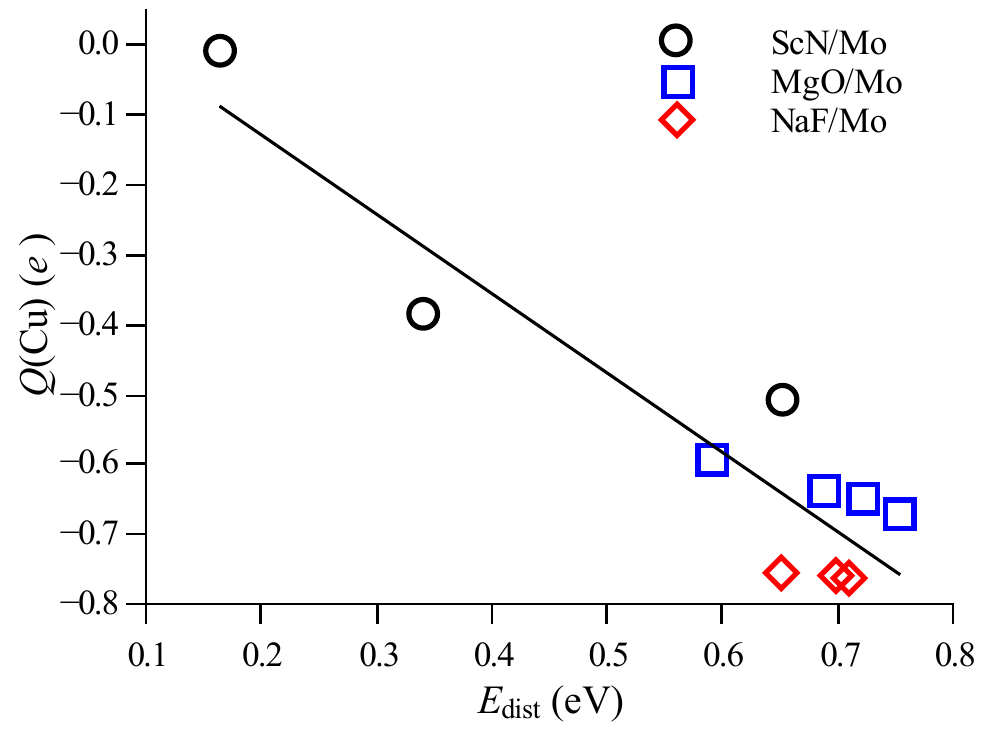}
\caption{\label{fig:Q_vs_Edist}(Color online) Charge of Cu adatom in Cu/AB/Mo 
versus energy of surface distortion, 
$E_\text{dist} = E(\text{[Cu]/AB/Mo}) - E(\text{AB/Mo})$.}
\end{figure}

Thus our results confirm the importance of surface relaxation for adatom 
charging. We also find it useful to study the characteristics of the AB/Mo 
systems having exactly the same surface distortion as the Cu/AB/Mo systems 
(i.e. when all A, B and Mo atoms are kept in the same positions as in Cu/AB/Mo, 
but Cu is removed). We denote such systems by [Cu]/AB/Mo. The energy needed to 
distort the [Cu]/AB/Mo surface is then 
$E_\text{dist} = E(\text{[Cu]/AB/Mo}) - E(\text{AB/Mo})$. $E_\text{dist}$ 
could be used as a measure of surface distortion. In Fig.~\ref{fig:Q_vs_Edist} 
we show the charge of Cu in Cu/AB/Mo versus distortion energy of [Cu]/AB/Mo, 
$E_\text{dist}$. There is a clear correlation between the energy of distortion 
and charge transferred to Cu, in fact, $E_\text{dist}$ and charge are linearly 
proportional to each other.

\subsection{\label{subsec:charge_localization}Distortion Induced Charge Localization on AB/Mo surface}

\begin{figure*}
\subfloat{\includegraphics[width = 0.32\textwidth]{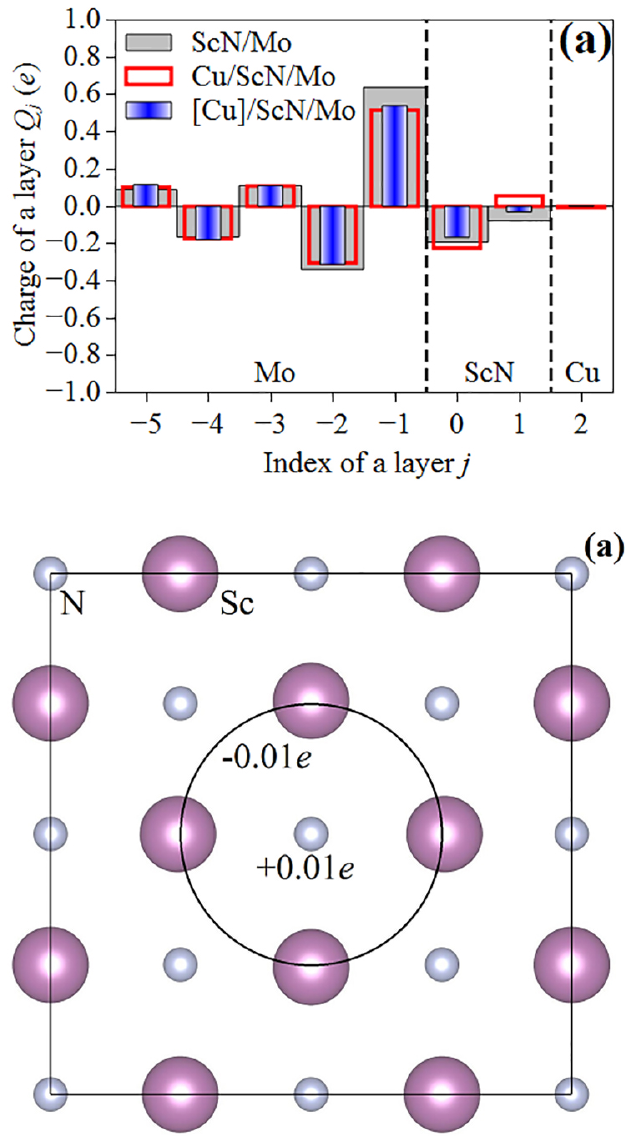}}\hfill
\subfloat{\includegraphics[width = 0.32\textwidth]{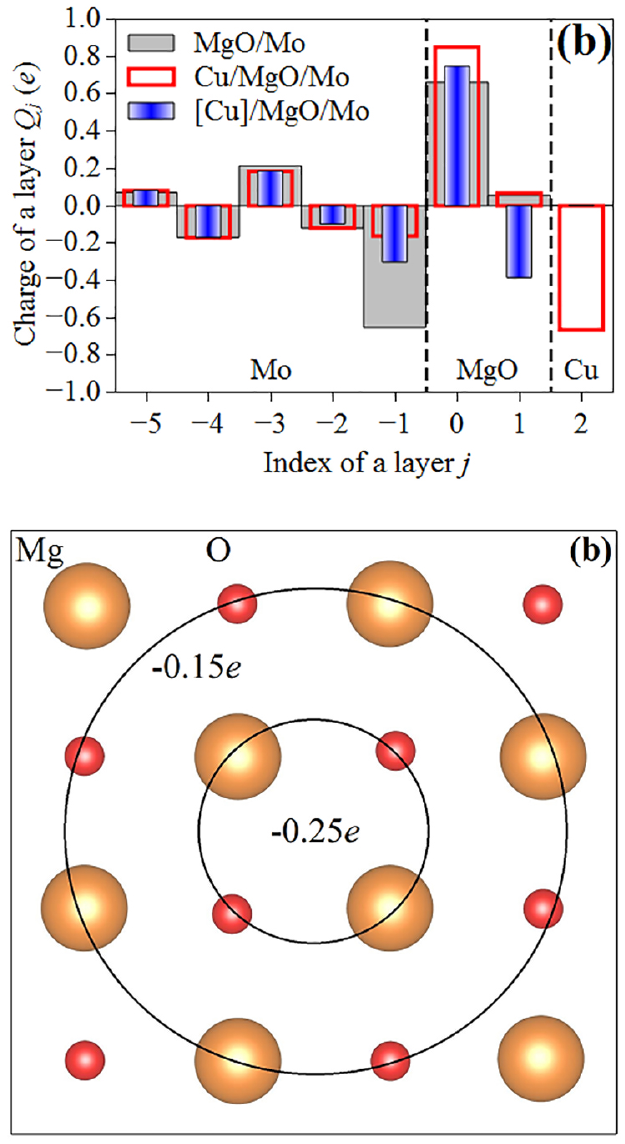}}\hfill
\subfloat{\includegraphics[width = 0.32\textwidth]{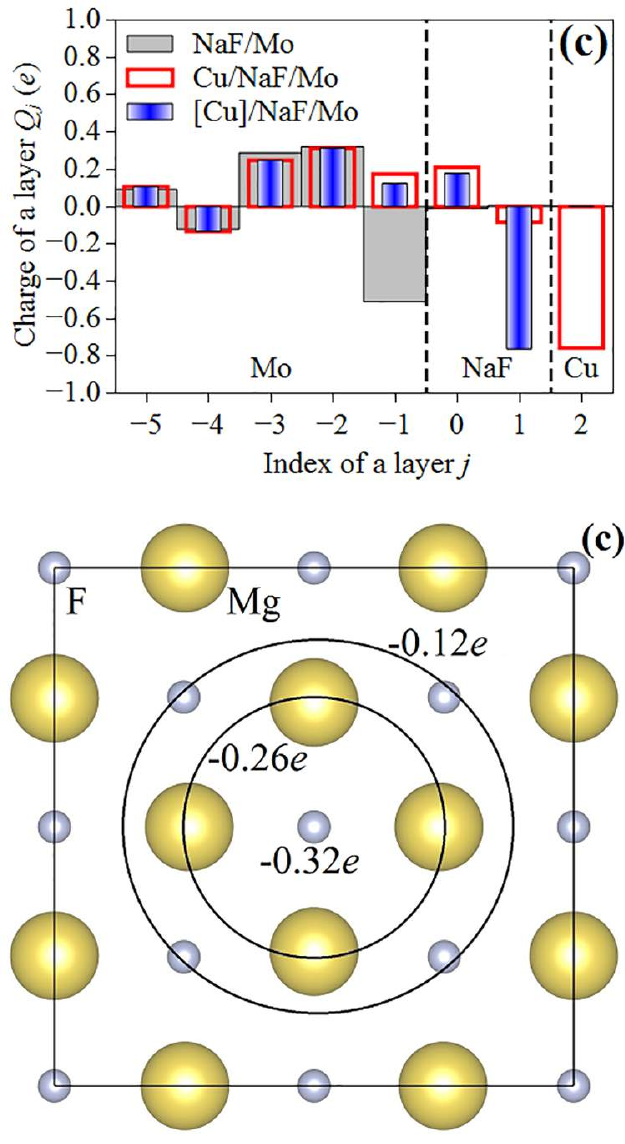}}
\caption{\label{fig:Q_ML}(Color online) Top panels: Comparison of charges of 
different layers of AB/Mo, Cu/AB/Mo and [Cu]/AB/Mo (has frozen geometry of 
relaxed Cu/AB/Mo, but Cu is absent). Bottom panels:  Accumulation of excess 
charge on specific atoms (or groups of atoms) in the top layer of AB, 
i.e. $Q_{\text{[Cu]/AB/Mo}} - Q_{\text{AB/Mo}}$. 
(a) N on-top site Cu/ScN/Mo, 
(b) hollow site Cu/MgO/Mo, and 
(c) F on-top site Cu/NaF/Mo.}
\end{figure*}

The charge accumulated by the Cu adatom comes from AB/Mo, but where is the 
origin of this charge? The analysis of the Bader charges of the substrate 
monolayers, calculated as described in Section~\ref{subsec:comp}, makes it 
possible to answer this question. Here we study the charge distribution in 
Cu/AB/Mo, [Cu]/AB/Mo and AB/Mo. In Fig.~\ref{fig:Q_ML} we show the layer 
charges for the three most stable configurations: N on-top site for Cu/ScN/Mo, 
hollow site for Cu/MgO/Mo and F on-top site for Cu/NaF/Mo.

Let us, first, compare Cu/AB/Mo and AB/Mo. Fig.~\ref{fig:Q_ML} (top panels) 
shows that for all Cu/AB/Mo the charge redistribution mostly occurs at the 
interface, while deeper Mo layers remain practically unaffected. For Cu 
adsorbed at the N on-top site, neither 
significant changes at the interface can be seen. For Cu adsorbed at the 
hollow site on MgO/Mo, where a significant charge transfer to Cu takes place, 
the charge of the adatom originates from the metal/oxide interface, mostly 
from the first Mo layer (Fig.~\ref{fig:Q_ML}b top panel).
These findings for 
Cu/MgO/Mo and Cu/NaF/Mo agree well with previous reports on Au/MgO/Mo and 
\ce{NO2}/MgO/Mo \cite{Frond08}.

Let us now consider distorted [Cu]/AB/Mo, which has the geometry of relaxed 
Cu/AB/Mo but without Cu. In terms of charge redistribution not much difference 
between Cu/ScN/Mo and [Cu]/ScN/Mo is observed (see Fig.~\ref{fig:Q_ML}a). For 
[Cu]/MgO/Mo and [Cu]/NaF/Mo, however, one can see a strong polarization of AB 
and Mo as a result of the distortion. Similar to Cu/MgO/Mo and Cu/NaF/Mo, the 
charge is accumulated in the top surface monolayer of MgO and NaF 
(see Fig.~\ref{fig:Q_ML}). The amount of charge accumulated in the surface 
layer is $-0.39e$ for MgO and $-0.77e$ for NaF. The origin of the charge is 
again the MgO/Mo (NaF/Mo) interface (see Fig.~\ref{fig:Q_ML}).

The Bader analysis of atomic charges shows that the excess charge, which for a 
particular atom we define as $Q_{\text{[Cu]/AB/Mo}} - Q_{\text{AB/Mo}}$, is 
spread over several atoms in the top layer (Fig.~\ref{fig:Q_ML}, bottom 
panels). In [Cu]/MgO/Mo most of the excess charge is accumulated in the 
surface layer around the hollow site and in the next coordination sphere. In 
[Cu]/NaF/Mo $-0.32e$ excess charge is localized at the F atom (below the Cu 
in Cu/NaF/Mo) and $-0.26e$ is localized at the four surrounding Na atoms. 
Thus, for both [Cu]/MgO/Mo and [Cu]/NaF/Mo the excess charge
appears to be spread over the distorted top layer and 
its larger part is located at the adsorption site. The amount of this excess 
charge accumulated in the distorted AB top layer is almost equal to the charge 
of Cu in the corresponding Cu/AB/Mo system. Thus, in both cases with and 
without Cu, the distortion leads to similar charge redistribution.
For more details on charge redistribution see Fig.~\ref{fig:spillover} in Appendix~A.

\subsection{\label{subsec:distortion_and_pumping}Charge Transfer From Metal Support to Adatom Through Thin Film: Cu/NaF/Mo case}

In this section we discuss the role of the thin film distortion in the charge 
transfer to the Cu adatom using the case of Cu adsorbed at the F on-top site 
on NaF/Mo (Cu/NaF/Mo), where charging is most pronounced.

\begin{figure*}
\subfloat{\includegraphics[width = 0.48\textwidth]{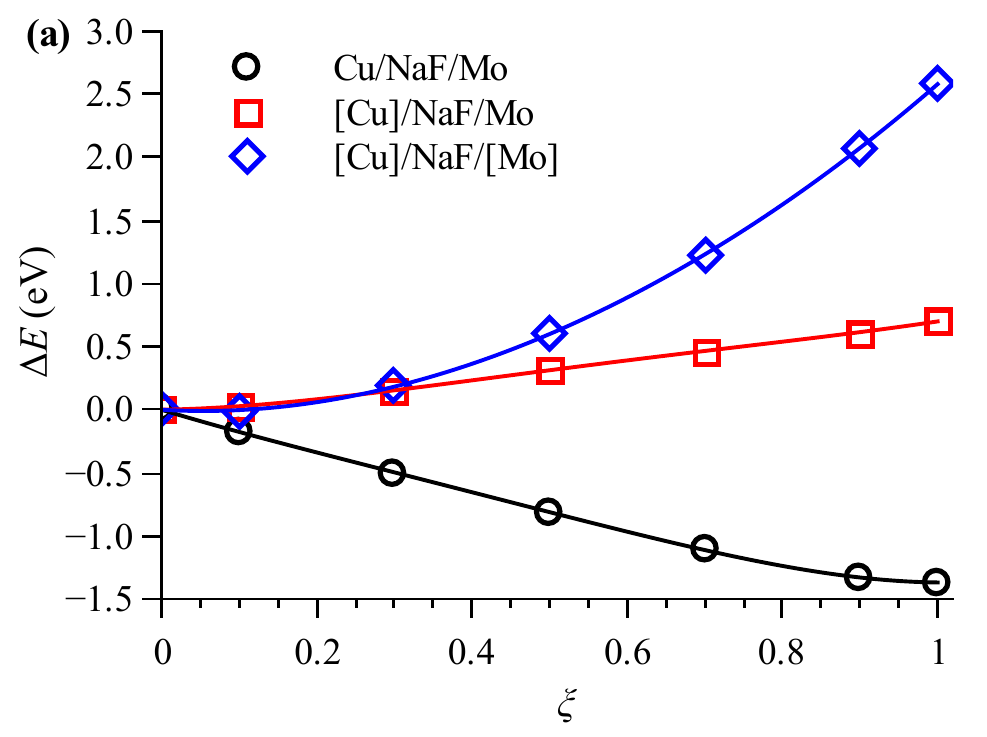}}\hfill
\subfloat{\includegraphics[width = 0.48\textwidth]{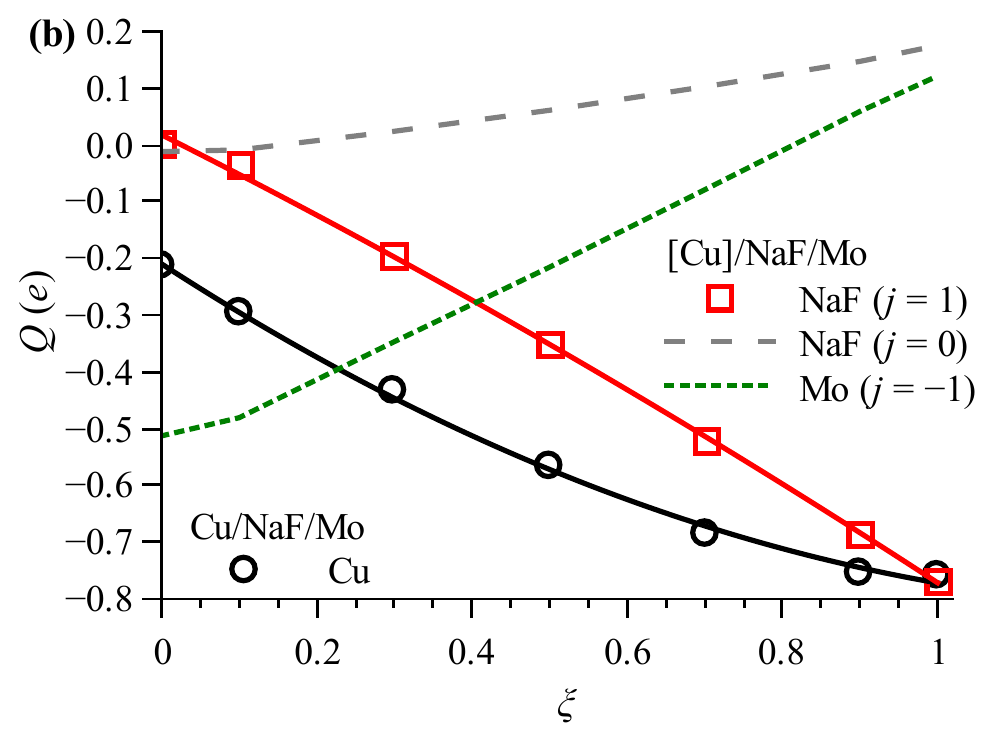}}\hfill
\caption{(Color online) Linear geometry transformation path
from $\xi = 0$ to $\xi = 1$ : 
(a) energy change along the path for Cu/NaF/Mo, [Cu]/NaF/Mo and [Cu]/NaF/[Mo], 
(b) charge of Cu adatom in Cu/NaF/Mo, and charges of NaF ($j = 1$) top surface 
layer, NaF ($j = 0$) interface layer and Mo ($j = -1$) interface layer in 
[Cu]/NaF/Mo (in the absence of Cu).}
\label{fig:E_and_Q_vs_xi}
\end{figure*}

To see how charge redistribution depends on the degree of deformation we 
approximated the deformation path of the surface with a linear geometry 
transformation. Let us denote the set of coordinates of all the Na and F atoms 
in NaF/Mo with $\bm{R}_0$ and those in Cu/NaF/Mo with 
$\bm{R}_1$, and introduce 
$\Delta \bm{R} = \bm{R}_1 - \bm{R}_0$.
Next, let us consider the linear path of geometry transformation from 
$\bm{R}_0$ to $\bm{R}_1$, which is given by 
$\bm{R} = \bm{R}_0 + \xi \Delta \bm{R}$, 
where $0 \leq \xi \leq 1$. For every value of $\xi$ the positions of Na and F 
atoms are defined and frozen, while Cu is allowed to relax.

Furthermore, we have considered the same linear transformation of NaF/Mo in 
the absence of Cu (the [Cu]/NaF/Mo case), as well as that of the NaF film in 
the absence of both Cu and Mo (the [Cu]/NaF/[Mo] case). Results are shown in 
Fig.~\ref{fig:E_and_Q_vs_xi}.

It is clear that in the case of Cu adsorption on NaF/Mo (Cu/NaF/Mo) the 
relaxation occurs spontaneously without any barrier. The charge of adatom 
increases almost linearly along the deformation path. Bader analysis shows 
that for both Cu/NaF/Mo and [Cu]/NaF/Mo the charges of NaF ($j = 0$, see 
Fig.~\ref{fig:slab}) and Mo ($j = 1$) interface layers monotonically decrease 
along the path (shown for [Cu]/NaF/Mo, see Fig.~\ref{fig:E_and_Q_vs_xi}b), 
thus charge is depleted from the NaF/Mo interface.

For Cu/NaF/Mo the gradually applied distortion yields gradual charging of Cu, 
while in the case of  [Cu]/NaF/Mo it provides monotonic charge accumulation in 
the NaF top layer ($j = 1$) (see Fig.~\ref{fig:E_and_Q_vs_xi}b). Moreover, we 
notice that along the path the values of the charge of Cu in Cu/NaF/Mo and the 
charge of NaF ($j = 1$) top layer in [Cu]/NaF/Mo are similar and become almost 
equal for $\xi = 1$.

Furthermore, our results indicate that the Na $3s$ and F $2p$ states participate
in the charge transfer from the NaF/Mo interface to the Cu adatom
(see Fig.~\ref{fig:Fermi_DOS} in Appendix B).

Our calculations show that charging requires distortion, i.e. cannot happen 
immediately (see Fig.~\ref{fig:E_and_Q_vs_xi}). There are indications that 
metal supported ultrathin films have different phonon structure compared to 
thick films and should be more flexible \cite{XPS_Freund}. Indeed, energy 
changes shown in Fig.~\ref{fig:E_and_Q_vs_xi}a demonstrate that it is much 
easier to deform [Cu]/NaF/Mo than [Cu]/NaF/[Mo]. The same distortion 
from $\xi = 0$  to $\xi = 1$ costs 0.70 eV for [Cu]/NaF/Mo and 2.58 eV for 
[Cu]/NaF/[Mo]. Moreover, the analysis of the curves show rather harmonic 
behavior for [Cu]/NaF/[Mo], while for [Cu]/NaF/Mo and Cu/NaF/Mo it is strongly 
anharmonic (see Fig.~\ref{fig:E_and_Q_vs_xi}a). Notice that with (or without) 
Cu adatom the energy change is almost linearly proportional to the accumulated 
charge. Hence, we attribute the NaF/Mo anharmonicity to the coupling of 
deformation with the charge transfer from the Mo/NaF interface to the NaF top 
layer (in [Cu]/NaF/Mo) or Cu adatom (in Cu/NaF/Mo).

\section{\label{sec:conclusions}Conclusions}
We have performed a systematic study of the AB/Mo and Cu/AB/Mo systems 
(AB = ScN, MgO, NaF). An enhancement of Cu adatom charging from ScN to MgO, 
and further to NaF is observed. The results suggest that charging is more 
pronounced when mixing between the states of the insulating film and metal 
substrate  as well as between those of an adatom and thin film is small.

The film/metal interface is the origin of adatom charge. The charge transfer 
is accompanied with a strong surface relaxation around the adatom and 
structural changes at the film/metal interface. Our results on Cu/AB/Mo show 
clear correlation between the amount of transferred charge and the degree of 
system distortion. Moreover, the results of the constrained deformation 
calculations indicate that distortion is decisive for the charge transfer and 
it will \textquotedblleft pump\textquotedblright\ the charge into the top 
surface layer irrespective of an adatom presence or absence on the surface.

We have found that the deformation of NaF/Mo film is essentially anharmonic, 
which we attribute to the coupling with the charge 
\textquotedblleft pumping\textquotedblright\ from metal/film interface to the 
film/vacuum interface or adatom. Also, it is much easier to distort NaF/Mo 
than NaF, therefore we expect that softening of some phonon modes of the metal 
supported films takes place.

\section{\label{sec:ack}Acknowledgements}
We would like to acknowledge Swedish Research Council (VR) and Swedish Energy 
Agency (STEM) for the support. We also thank the Swedish National 
Infrastructure for Computing (SNIC) for provided computational resources. 
P.A.\v{Z}. acknowledges A. V. Ruban for interesting discussions.
P.A.\v{Z}. also acknowledges VESTA~\cite{VESTA} software developers.

\appendix
\setcounter{secnumdepth}{0}
\section{\label{AppA}Appendix A: Spatial Charge Distribution}

\begin{figure}[!h]
\includegraphics[width = 3.375in]{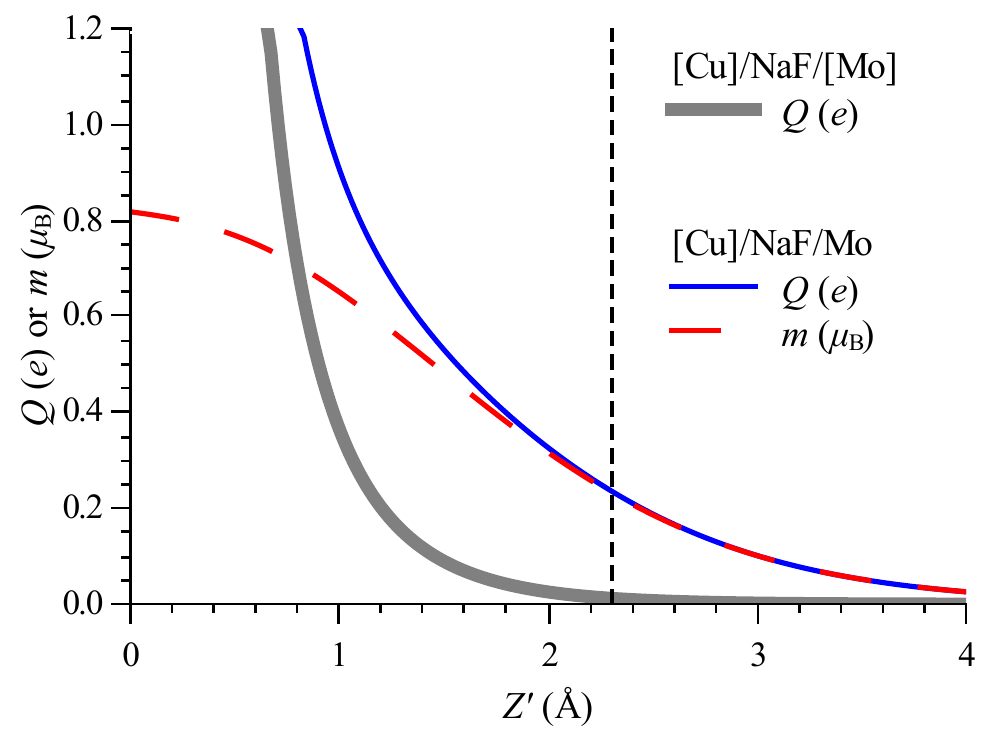}
\caption{\label{fig:spillover} The comparison of [Cu]/NaF/[Mo] with
[Cu]/NaF/Mo showing charge spillover for the latter. Charge or magnetization 
integrated between $z = Z'$ plane and the middle of the vacuum 
gap. $Z' = 0$ corresponds to the $z$ coordinate of [Cu]/NaF/Mo surface 
termination.}
\end{figure}

The spatial distribution of charge accumulated in [Cu]/NaF/Mo was examined 
further.
In Fig.~\ref{fig:spillover} we show charge density for [Cu]/NaF/Mo 
and [Cu]/NaF/[Mo] integrated along the $Z$ axis.
Let us recall, that [Cu]/NaF/Mo (also [Cu]/NaF/[Mo])
have the same geometry as Cu/NaF/Mo, but Cu (respectively Cu and Mo) are absent.
One can see that in the 
[Cu]/NaF/Mo case charge is more spread in the $Z$ direction compared to the 
case when both Cu adatom and Mo support are removed ([Cu]/NaF/[Mo]).
In particular, for [Cu]/NaF/Mo approximately $0.2e$ are found
above the position of Cu in Cu/AB/Mo 
(note, there is no Cu in [Cu]/NaF/Mo).
The same conclusion follows from the 
integrated magnetization curve, which agrees well with the integrated charge 
curve (Fig.~\ref{fig:spillover}).

\appendix
\section{\label{AppB}Appendix B: DOS of Na and F during deformation}

\begin{figure*}[!t]
\subfloat{\includegraphics[width = 0.32\textwidth]{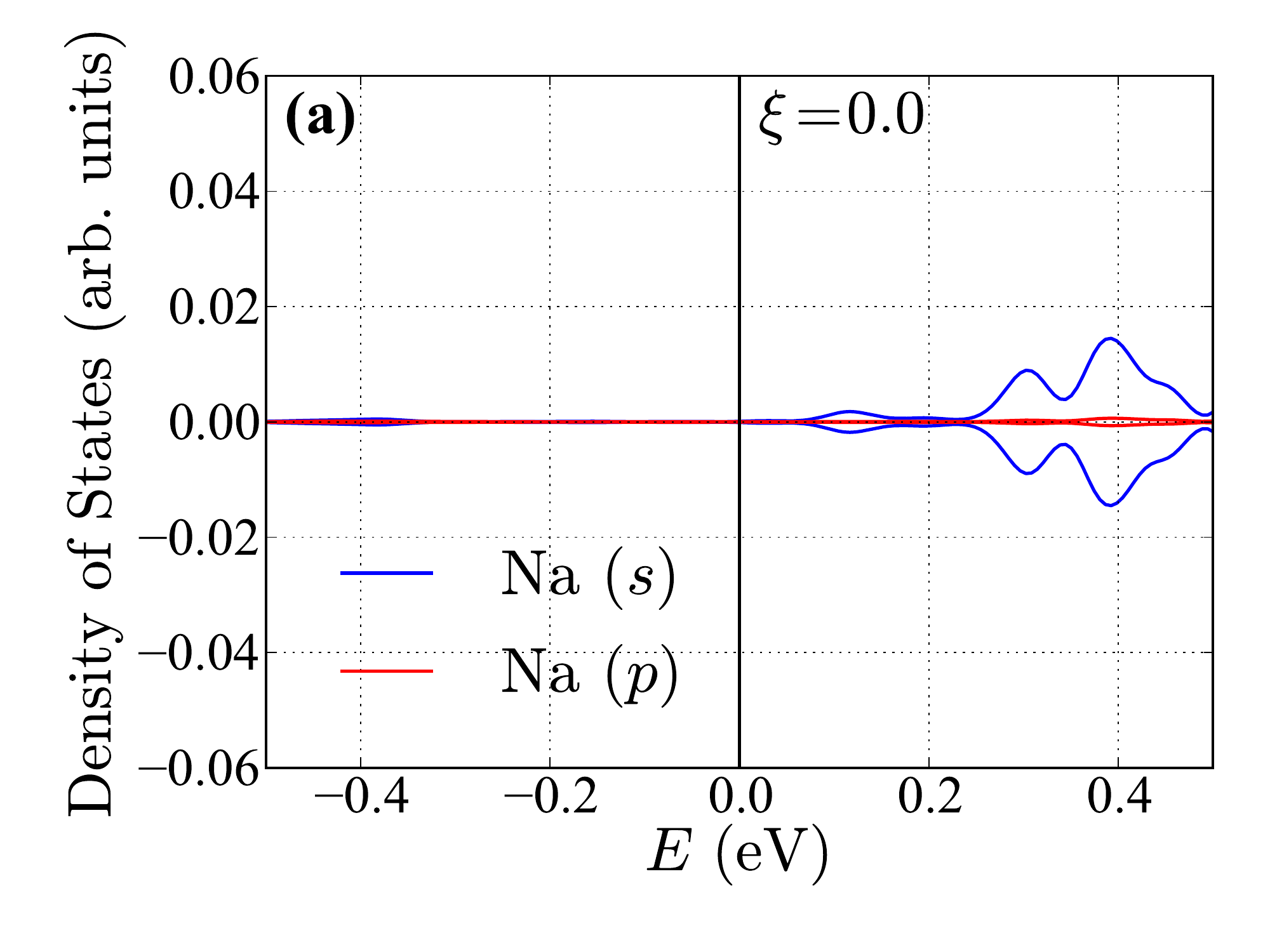}}\hfill
\subfloat{\includegraphics[width = 0.32\textwidth]{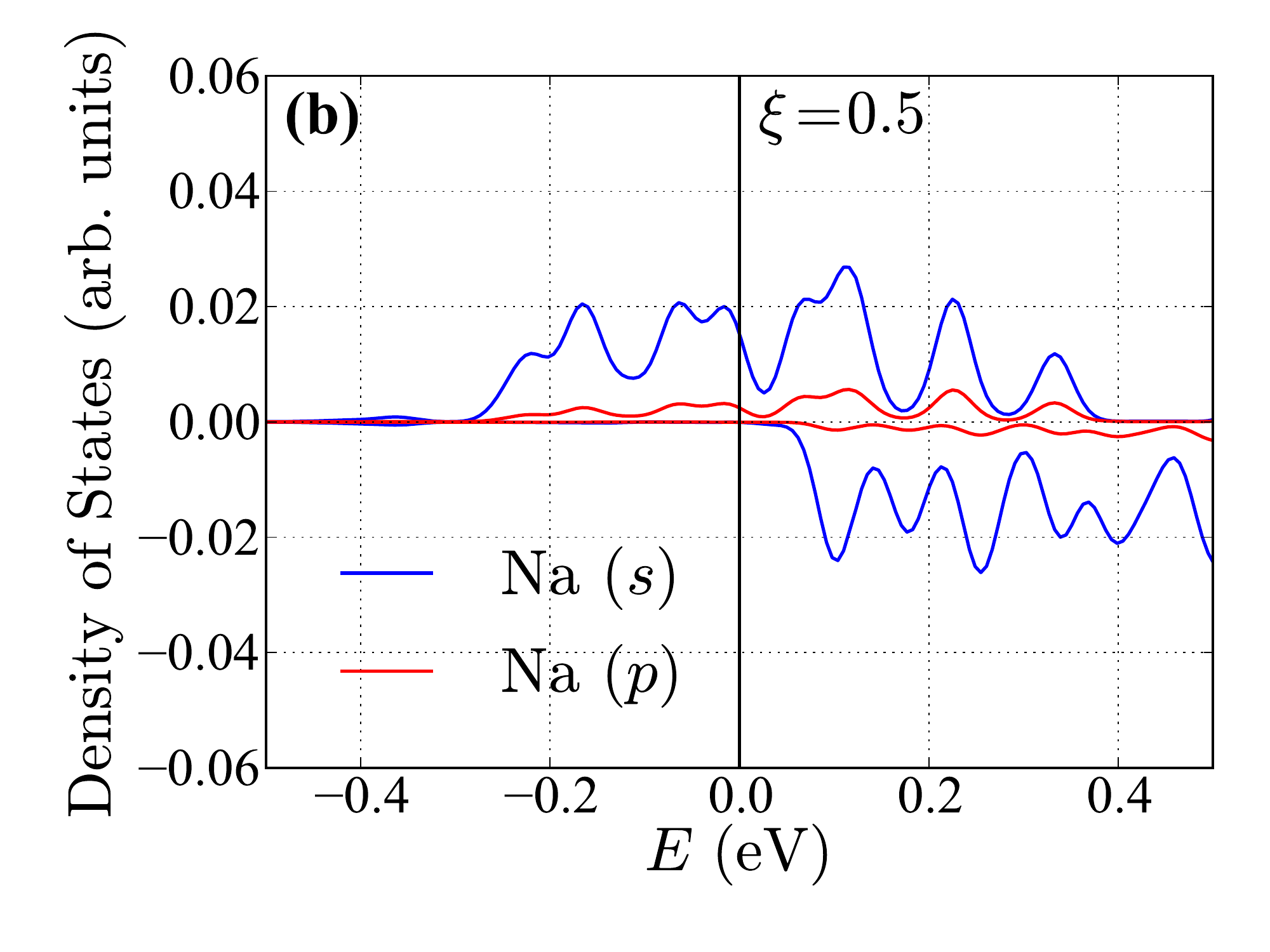}}\hfill
\subfloat{\includegraphics[width = 0.32\textwidth]{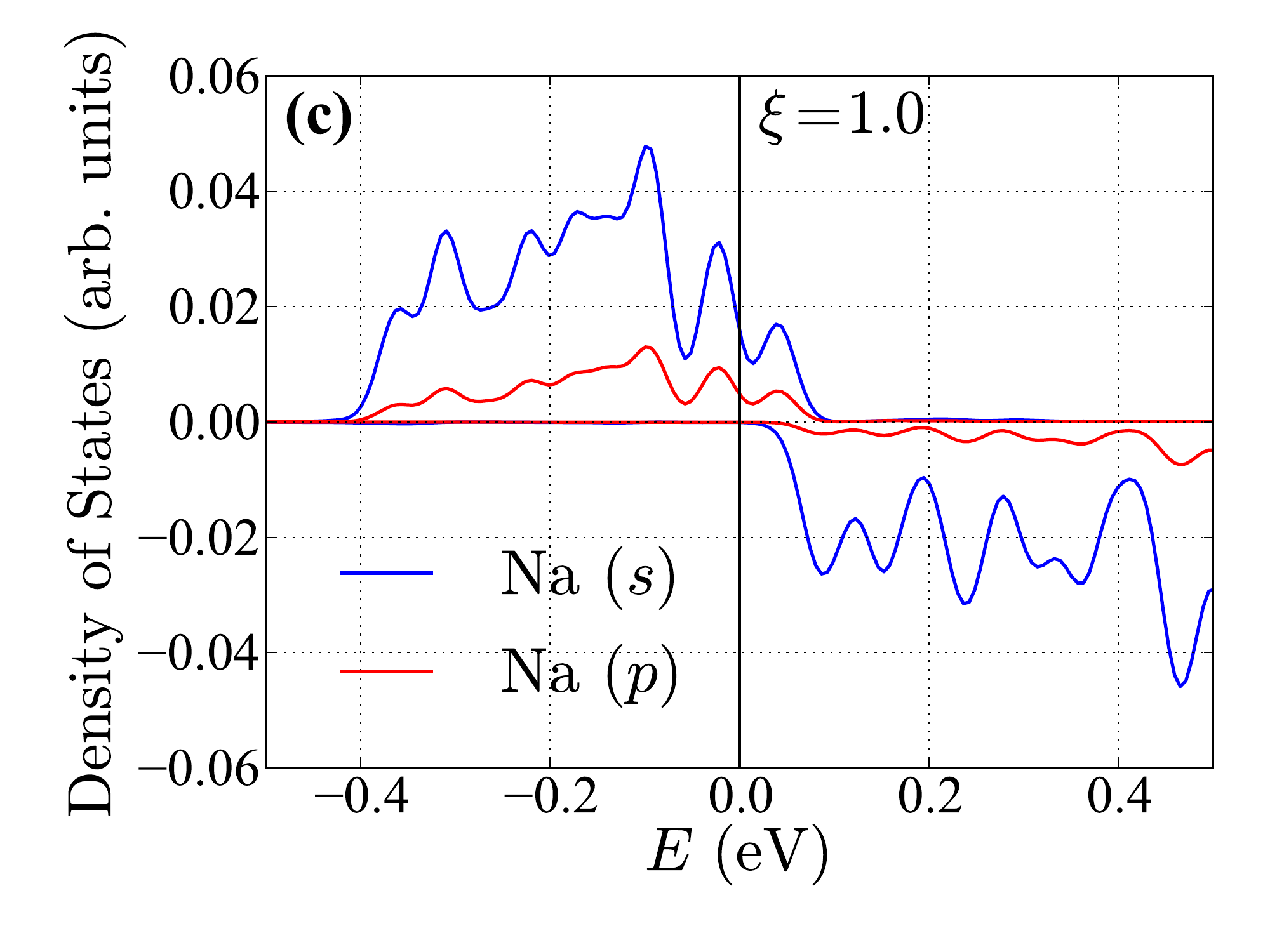}}\hfill \\
\subfloat{\includegraphics[width = 0.32\textwidth]{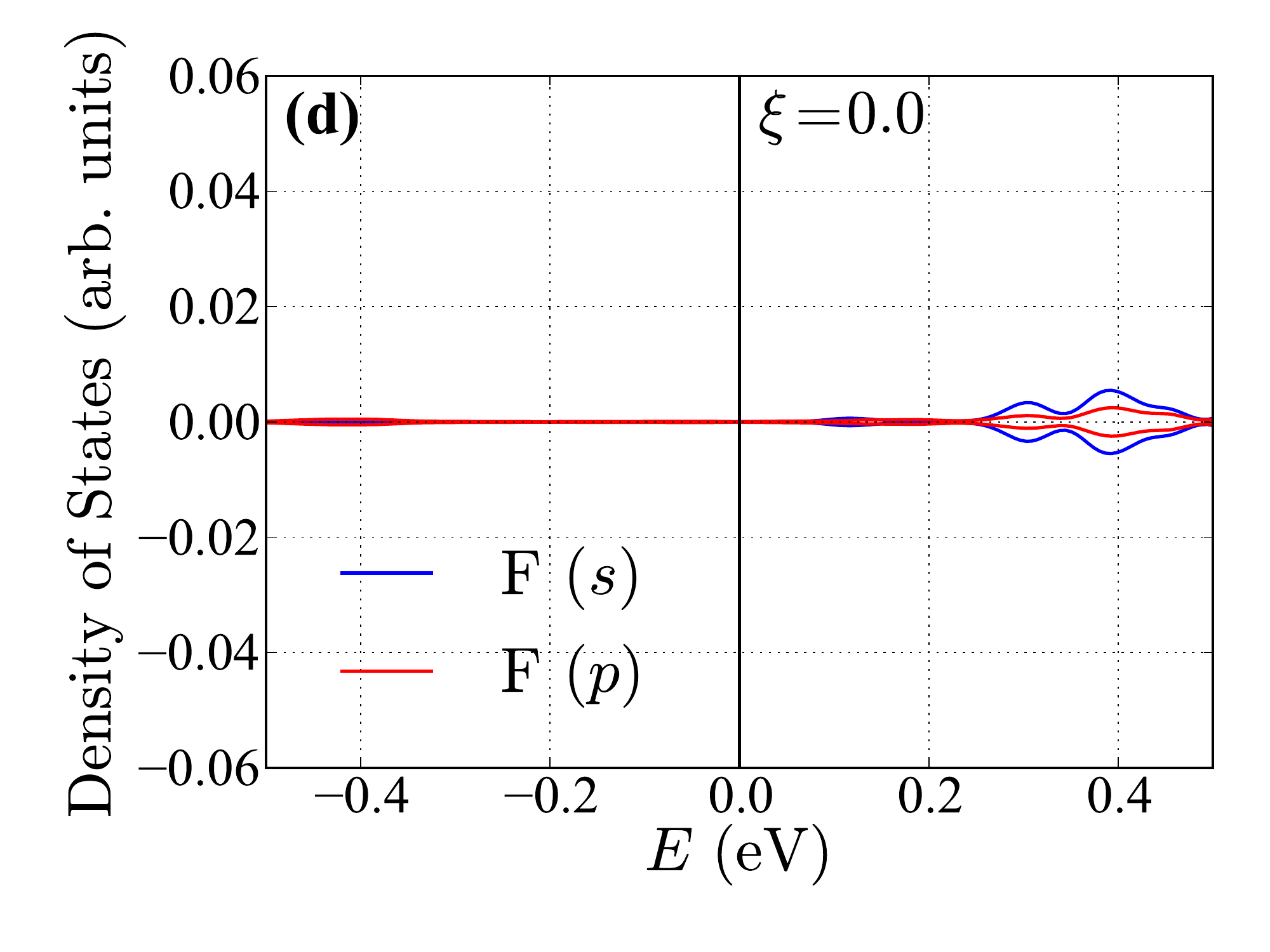}}\hfill
\subfloat{\includegraphics[width = 0.32\textwidth]{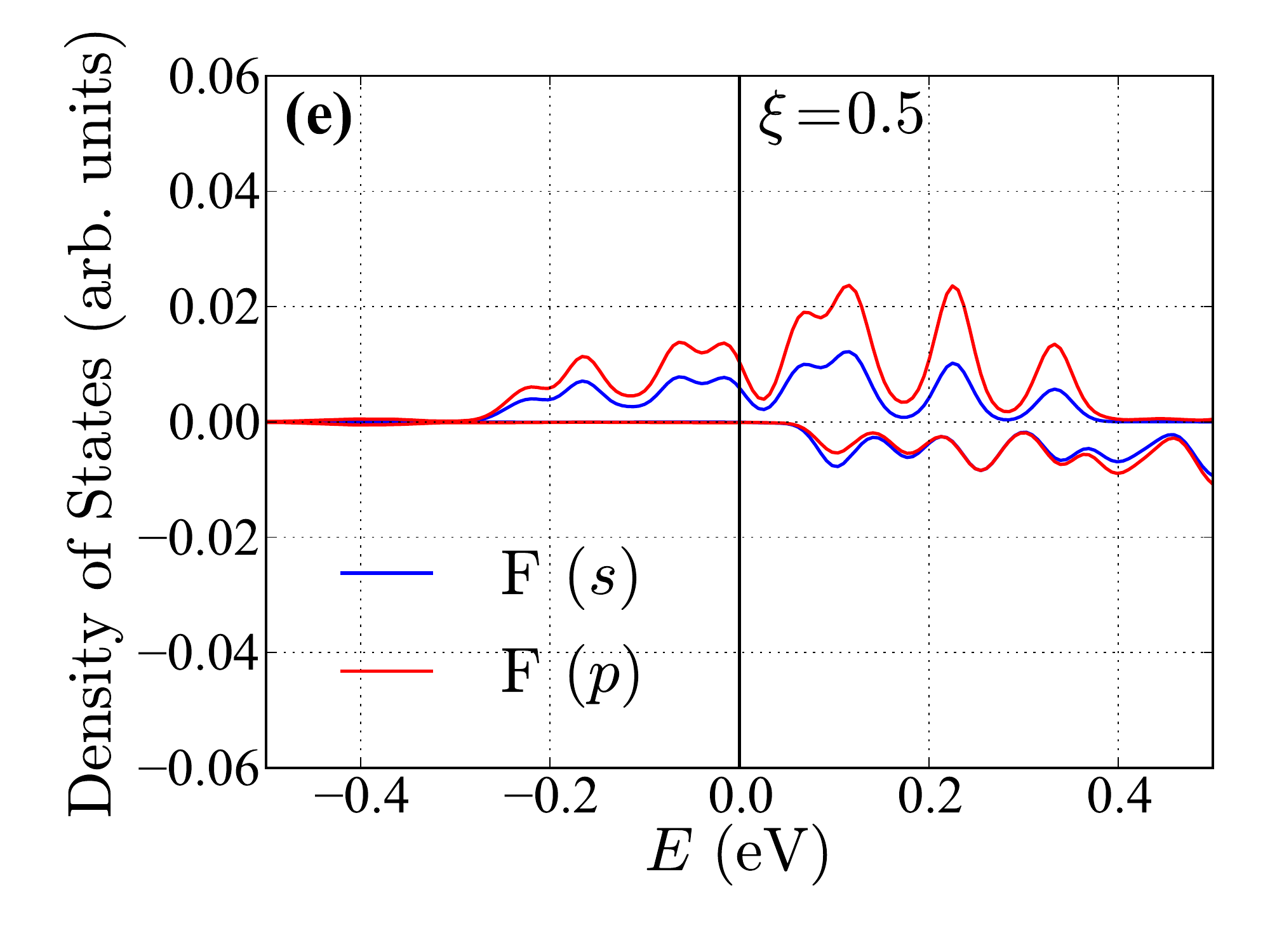}}\hfill
\subfloat{\includegraphics[width = 0.32\textwidth]{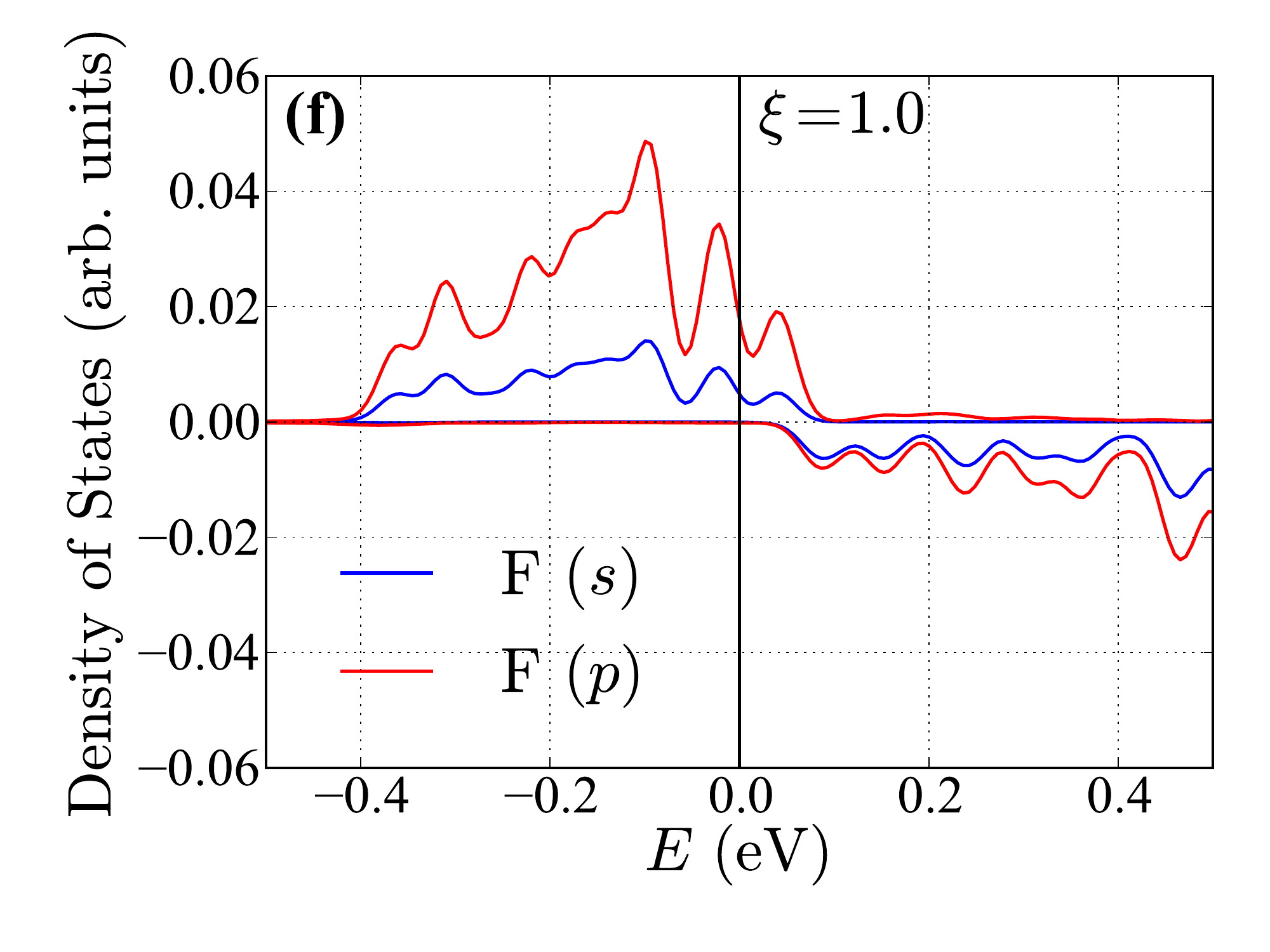}}
\caption{\label{fig:Fermi_DOS} [Cu]/NaF/Mo: partial density of 
states near the Fermi level (set to zero): (a), (b), and (c) Na atom (which is 
near Cu adatom in Cu/NaF/Mo), (d), (e), and (f) F atom (which is below Cu 
adatom in Cu/NaF/Mo). In (a) and (d) $\xi = 0.0$, in (b) and (e) $\xi = 0.5$, 
in (c) and (f) $\xi = 1.0$.}
\end{figure*}

Here we report the details behind the distortion of [Cu]/NaF/Mo from
$\xi = 0$ to $\xi = 1$ (see the main text for explanations).
In Fig.~\ref{fig:Fermi_DOS} we show the evolution of the density of 
states near the Fermi level for [Cu]/NaF/Mo, namely, for the F atom underneath 
Cu in Cu/NaF/Mo and the nearest Na atoms in the top NaF layer. According to 
the Bader analysis these atoms accumulate the charge, while Mo atoms in the 
interface layer lose it (not shown) that agrees well with the evolution of the 
density of states. The number of the F states below the Fermi level (mainly 
the $2p_z$ states) and Na states (mainly the $3s$ states) gradually increases 
along the deformation path (see Fig.~\ref{fig:Fermi_DOS}), while the number of 
the Mo states decreases (not shown). As a result of the charge transfer the 
system becomes spin polarized, near the Fermi level the spin-down states are 
unoccupied (see Fig.~\ref{fig:Fermi_DOS}). Therefore, our findings 
indicate that the F $2p$ and Na $3s$ states participate in the charge 
transfer to Cu adatom.

\clearpage

\newpage

\newpage

\newpage

%\bibliography{References}
%\bibliographystyle{plainnat}


\end{document}